\documentclass[journal]{IEEEtran}
\usepackage[compress]{cite}
\usepackage[utf8]{inputenc}
\usepackage{amsmath}
\usepackage{amsfonts}
\usepackage{amssymb}
\usepackage{graphicx}
\usepackage{algorithmic}
\usepackage{algorithm}
\usepackage{color}
\usepackage{subfigure}
\usepackage{bm}
\usepackage{algorithm}
\usepackage{algorithmic}
\usepackage{sidecap}
\usepackage{booktabs}
\usepackage{tikz}

\newcommand{\argmax}{\operatornamewithlimits{arg\ max}}

\definecolor{darkgreen}{rgb}{0,0.6,0.2}

\begin{document}

\title{Determination of the edge of criticality in echo state networks through Fisher information maximization}

\author{Lorenzo Livi,~\IEEEmembership{Member,~IEEE},
        Filippo Maria Bianchi,
       and Cesare Alippi,~\IEEEmembership{Fellow,~IEEE}
\thanks{Manuscript received ; revised .}
\thanks{Lorenzo Livi and Cesare Alippi are with the Dept. of Electronics, Information, and Bioengineering, Politecnico di Milano, Italy and Faculty of Informatics, Universit\`a della Svizzera Italiana, Lugano, Switzerland (e-mail: \{lorenzo.livi, cesare.alippi\}@polimi.it).}
\thanks{Filippo Maria Bianchi is with the Machine Learning Group, Dept. of Physics and Technology, University of Troms\o{}, the Arctic University of Norway (e-mail: filippo.m.bianchi@uit.no).}}

\maketitle

\begin{abstract}
It is a widely accepted fact that the computational capability of recurrent neural networks is maximized on the so-called ``edge of criticality''.
Once the network operates in this configuration, it performs efficiently on a specific application both in terms of (i) low prediction error and (ii) high short-term memory capacity. Since the behavior of recurrent networks is strongly influenced by the particular input signal driving the dynamics, a universal, application-independent method for determining the edge of criticality is still missing.
In this paper, we aim at addressing this issue by proposing a theoretically motivated, unsupervised method based on Fisher information for determining the edge of criticality in recurrent neural networks.
It is proven that Fisher information is maximized for (finite-size) systems operating in such critical regions.
However, Fisher information is notoriously difficult to compute and either requires the probability density function or the conditional dependence of the system states with respect to the model parameters.
The paper takes advantage of a recently-developed non-parametric estimator of the Fisher information matrix and provides a method to determine the critical region of echo state networks, a particular class of recurrent networks.
The considered control parameters, which indirectly affect the echo state network performance, are explored to identify those configurations lying on the edge of criticality and, as such, maximizing Fisher information and computational performance.
Experimental results on benchmarks and real-world data demonstrate the effectiveness of the proposed method.
\end{abstract}
\begin{IEEEkeywords}
Edge of criticality; Echo state network; Fisher information; Non-parametric estimation.
\end{IEEEkeywords}

 \section*{Nomenclature}
 
 \begin{center}\small
 \begin{tabular}{ll}
 ARIMA & Autoregressive integrated moving average \\
 ESN	 & Echo state network \\
 FIM	 & Fisher information matrix \\
 GA & Genetic algorithm \\
 MC & Memory capacity \\
 MLLE & Maximum local Lyapunov exponent \\
 MST	 & Minimum spanning tree \\
 mSVJ & Minimum singular value of Jacobian \\
 NARMA & Non-linear autoregressive moving average \\
 NRMSE & Normalized root mean square error \\
 PD & Positive semidefinite \\
 PDF	 & Probability density function \\
 RNN	 & Recurrent neural network \\
 SVR	 & Support vector regression \\
 $\phi$ & Region in parameter space where FIM is maximized \\
 $\lambda$ & Region in parameter space where MLLE crosses zero \\
 $\eta$ & Region in parameter space where mSVJ is maximized \\
 $\gamma$ & Prediction accuracy \\
 \end{tabular}
 \end{center}

\section{Introduction}

A Recurrent Neural Network (RNN) can approximate any dynamic system under mild hypotheses (see \cite{maass2007computational} and references therein). However, RNNs are difficult to train \cite{pascanu2012difficulty} and the interpretability of their modus operandi is still object of study \cite{sussillo2013opening,bianchi2016investigating}.
Interestingly, RNNs can generate complex dynamics even characterized by sharp transitions permitting them to commute between ordered and chaotic regimes.
In fact, experimental results on a multitude of application contexts suggest that RNNs achieve the highest information processing capability exactly when configured on the edge of this transition, resulting in high memory capacity (storage of past inputs) and good performance on the modeling/prediction task at hand (low prediction errors) \cite{legenstein2007edge,bertschinger2004real,toyoizumi2011beyond,rajan2010stimulus,PhysRevLett.114.088101,PhysRevLett.110.118101}.
Therefore, in order to determine such ``critical'' network configurations, RNNs require fine tuning of their controlling parameters.
This general behavior is in agreement with the widely-discussed ``criticality hypothesis'' associated with the functioning of many biological (complex) systems \cite{scheffer2009early,tkavcik2014information,hidalgo2014information,grigolini2015emergence,mora2011biological,roli2015dynamical}, including the brain \cite{torres2015brain,de2014criticality,hesse2014self,mora2015dynamical,massobrio2015criticality,tkavcik2015thermodynamics,moretti2013griffiths}.
In fact, it was noted, e.g., see \cite{mora2011biological}, that such complex systems tend to self-organize so as to operate in a critical regime. This still controversial issue has been supported by experiments showing that, in such a regime, systems are highly responsive to external stimuli and hence capable of introducing any dynamics as requested by the specific task \cite{mora2011biological}.
Investigating whether a given complex system operates more efficiently in the critical regime or not requires, at first, theoretically sound methods for detecting the onset of criticality \cite{scheffer2012anticipating}.

Determination of system configurations lying on the edge of criticality can be then carried out by means of appropriate sensitivity analyses (on the edge of criticality the sensitivity diverges, being the separation between ordered and chaotic regimes).
In this direction, Fisher information, and its multivariate extension called Fisher information matrix (FIM) \cite{zegers2015fisher,toyoizumi2006fisher,beck2011insights,transtrum2015perspective}, provide a way to quantify the sensitivity of a (parametrized) probability density function with respect to its control parameters.
Fisher information is tightly linked with statistical mechanics and, in particular, with the field of (continuous) phase transitions.
In fact, as shown in \cite{prokopenko2011relating}, it is possible to provide a thermodynamic interpretation of Fisher information in terms of rate of change of the order parameter, quantities used to discriminate the different phases of a system.
This fact provides an important bridge between the concept of criticality and statistical modeling of complex systems.
It emerges that the critical phase of a thermodynamic system can be mathematically described as that region of the parameter space where the order parameters vanish and their derivatives diverge. This implies that, on the critical region, Fisher information diverges as well, hence providing a quantitative, well-justified tool for detecting the onset of criticality in both theoretical models and computational simulations \cite{6792593}.
Nonetheless, Fisher information is notoriously difficult to compute and, in principle, it requires the analytical knowledge of the parametrized probability density function describing the system behavior.

The designer could consider directly the network weights and drive them towards the edge of criticality through a learning mechanism. Even though this problem is still open at current state of research, what we propose here strongly goes in this direction by accounting a special class of RNNs called Echo State Networks (ESNs) \cite{lukovsevivcius2009reservoir}.
Although ESNs are typically randomly initialized, the network designer has access to a set of hyperparameters, which have an indirect effect (when considering inputs) on the resulting ESN dynamics and their related computational capability.
We define the hyperparameter configurations that bring an ESN in a state where prediction accuracy and memory capacity are maximized as the critical region (or equivalently, edge of criticality).
Here we show that the FIM can be used to determine the onset of criticality for a network designed to solve a particular application.
Notably, we provide an unsupervised algorithm that exploits the determinant of the FIM in order to determine the edge of criticality.
Since the proposed algorithm is unsupervised, it does not depend on the particular model and related training mechanism adopted for the readout.
This feature becomes particularly relevant when the readout layer is implemented by means of non-linear models, such as feed-forward neural networks or kernel-based support vector regression, which require a long training time.
In the proposed algorithm, we use a non-parametric FIM estimator \cite{6975144} that allows us to overcome some of the difficulties that arise when adopting a model-based approach to compute the FIM (e.g., availability of the analytical model ruling the system).
Additionally, in order to robustly estimate the FIM, we follow an ensemble approach and perform a number of independent trials.

RNNs, as well as ESNs, are driven by inputs. Therefore, their dynamics and related computational capability depend on the type of input signal driving the network.
During the last decade, many solutions have been proposed to characterize the input-driven dynamics of the network and perform related tuning of the (hyper-)parameters \cite{obst2014guided}.
Among the many contributions, we can cite approaches based on mean-field approximation of the neuron activations \cite{massar2013mean}, information-theoretic methods inspired by the concept of intrinsic plasticity (based on the maximum entropy principle) borrowed from neuroscience \cite{ozturk2007analysis,schrauwen2008improving}, and methods for characterizing the onset of criticality with measures of (directional) information transfer and information storage \cite{boedecker2012information}, together with related self-organized adaptation mechanisms \cite{obst2010improving}.

To the best of our knowledge, FIM and related thermodynamic interpretations have not been considered yet to study the issue of criticality in ESNs.
We stress that, in principle, our method can be extended to account for additional (hyper)parameters, such as feedback scaling and percentage of noise in state update \cite{jaeger2002adaptive}.
Finally, it is worth noticing that, as a consequence of the theoretical framework adopted here, we implicitly assume that the critical phase of ESNs can be described by a continuous phase transition. This assumption is highly justifiable, since a system can operate in a critical regime only if such a transition is continuous.

The novelty of our contribution can be summarized as:
\begin{itemize}
 \item An unsupervised learning method that, by exploiting the information coming from the neuron activations only, permits to identify the edge of crticality.
 Since no assumption regarding the mathematical model of the (input-driven) dynamic system is made, the method can handle any type of applications;
 \item The proposed method is independent of the particular reservoir topology, since it is conceived to determine the critical ESN (hyper-)parameters. This allows the network designer to instantiate a specific architecture based on problem-dependent design choices;
 \item The envisaged non-parametric FIM estimator \cite{6975144} operates directly on data/observations: as such, there is no need to estimate the high-dimensional densities underlying the neuron activations. As a consequence, the number of reservoir neurons does not pose a serious technical issue from the estimation viewpoint and therefore it can be chosen by the network designer according to application requirements;
 \item The FIM estimator can be implemented in two different ways, one of which requires elaboration in order to properly define the related optimization problem.
In this paper, we propose our own formulation for the constraints defining such an optimization problem -- see Appendix \ref{sec:constraint_implementation} for details.
\end{itemize}

The remainder of this paper is structured as follows.
In Section \ref{sec:ESN} we introduce ESNs and the related considerations on the characterization of the dynamics.
Section \ref{sec:FIM_estimator} introduces Fisher information matrix and the adopted non-parametric estimator.
In Section \ref{sec:proposed_method}, we present the method that we propose for determining the ESN hyperparameters.
In order to support our methodological developments, Section \ref{sec:exps} presents experimental results performed on both well-known benchmarks and a real-world application involving the prediction of telephone call loads \cite{bianchi2015prediction}.
Conclusions and future research directions follow in Section \ref{sec:conclusions}.

\section{Echo state networks}
\label{sec:ESN}

\begin{figure}[htp!]
    \centering
    \includegraphics[width=0.55\columnwidth]{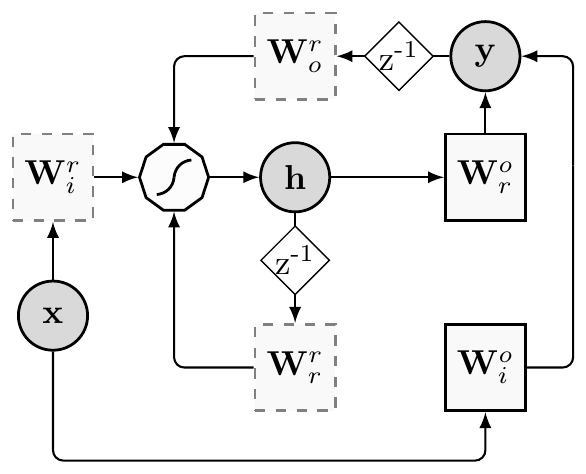}
    \caption{Schematic depiction of an ESN. The circles represent input $\mathbf{x}$, state, $\mathbf{h}$, and output, $\mathbf{y}$, respectively. Solid squares $\mathbf{W}_{r}^{o}$ and $\mathbf{W}_{i}^{o}$, are the trainable matrices, respectively, of the readout, while dashed squares, $\mathbf{W}_{r}^{r}$, $\mathbf{W}_{o}^{r}$, and $\mathbf{W}_{i}^{r}$, are randomly initialized matrices. The polygon represents the non-linear transformation performed by neurons and $\text{z}^{\text{-1}}$ is the time lag operator.}
    \label{fig:esn}
\end{figure}

ESNs \cite{lukovsevivcius2009reservoir} consist of a large recurrent layer of non-linear units with randomly generated weights and a linear, memory-less readout layer that is trainable by means of a simple regularized least-square optimization.
The recurrent layer acts as a non-linear kernel \cite{hermans2012recurrent}, mapping the input to a high-dimensional space.
A visual representation of the ESN architecture is reported in Fig. \ref{fig:esn}.

The equations describing the ESN state-update and output are, respectively, defined as
\begin{align}
\label{eq:state_update}
\mathbf{h}[k] =& \phi(\mathbf{W}_{r}^{r}\mathbf{h}[k-1] + \mathbf{W}_{i}^{r}\mathbf{x}[k] + \mathbf{W}_{o}^{r}\mathbf{y}[k-1]),\\
\label{eq:esn_output}
\mathbf{y}[k] =& \mathbf{W}_{i}^{o}\mathbf{x}[k] + \mathbf{W}_{r}^{o}\mathbf{h}[k].
\end{align}

The reservoir contains $N_r$ neurons characterized by a transfer/activation function $\phi(\cdot)$, which is typically implemented as a hyperbolic tangent (\emph{tanh}) function.
At time instant $k$, the network is driven by the input signal $\mathbf{x}[k]\in \mathbb{R}^{N_i}$ and produces the output $\mathbf{y}[k] \in \mathbb{R}^{N_o}$, being $N_i$ and $N_o$ the dimensionality of inputs and outputs, respectively.
The vector $\mathbf{h}[k]$ contains $N_r$ components and describes the ESN (instantaneous) state.
The weight matrices $\mathbf{W}_r^r \in \mathbb{R}^{N_r \times N_r}$ (reservoir connections), $\mathbf{W}_i^r \in \mathbb{R}^{N_i \times N_r}$ (input-to-reservoir), and $\mathbf{W}_o^r \in \mathbb{R}^{N_o \times N_r}$ (output-to-reservoir feedback) contain real values in the $[-1, 1]$ interval distributed according to a uniform distribution, but additional options have been explored in the recent literature \cite{rodan2011minimum,appeltant2011information}.
$\mathbf{W}_{i}^{o}$ and $\mathbf{W}_{r}^{o}$, instead, are optimized for the task under consideration, usually by means of a (regularized) linear regression algorithm.
Here, for the sake of brevity we do not report the expressions describing training (regularized linear regression of ESN readout) and refer the reader to \cite{lukovsevivcius2009reservoir} for details. In fact, the proposed method for finding the hyperparameters is completely unsupervised and, hence, independent from the readout training.

The behavior of a given network can be controlled by tuning a set of scalar hyperparameters. Usually, the designer considers $\theta_{IS}$, the scaling of the input weights $\mathbf{W}_i^r$, hence affecting the non-linearity introduced by the neurons; $\theta_{SR}$, scaling of the spectral radius of $\mathbf{W}_r^r$, which influences both stability and computational capability of the network by shifting the transfer function poles \cite{ozturk2007analysis}; $\theta_{RC}$, which determines the sparsity of connectivity in $\mathbf{W}_r^r$, i.e., the number of weights set to 0; $\theta_{FB}$, which affects $\mathbf{W}_{r}^{o}$, that is, the importance of output feedback connections.
In this study, we set $\theta_{FB}=0$ with a consequent simplification of ESN state-update (\ref{eq:state_update}):
\begin{equation}
\label{eq:esn_state_nooutputfeedback}
\mathbf{h}[k] = \phi(\mathbf{W}_{r}^{r}\mathbf{h}[k-1] + \theta_{IS}\mathbf{W}_{i}^{r}\mathbf{x}[k]),
\end{equation}
where $\mathbf{W}_{r}^{r} = \theta_{SR} \mathbf{W}_{r}^{r} / \rho(\mathbf{W}_{r}^{r})$, being $\rho(\mathbf{W}_{r}^{r})$ the spectral radius of $\mathbf{W}_{r}^{r}$.
$\theta_{SR}$, $\theta_{IS}$, and $\theta_{RC}$ are hyperparameters typically tuned through cross-validation to find the best-performing configuration for the task at hand.
In this paper, we study how to set these three hyperparameters through an unsupervised approach. However, we stress that the proposed methodology is applicable to any number of hyperparameters.

In order to guarantee asymptotic stability, ESNs must satisfy the so-called echo state property \cite{caluwaerts2013spectral,1629106,manjunath2013echo,yildiz2012re}, which requires the reservoir exhibiting short-term memory (exponential fading) \cite{tivno2013short,ganguli2008memory}. Recently, in \cite{mayer2015input} the author investigated the effects of criticality in ESN memory, showing that, under certain conditions, the echo state property can still be verified even if the memory vanishes more slowly (i.e., following a power-law function).

The stability margin of a network can be assessed in practice by analyzing the Jacobian matrix of the reservoir state update (\ref{eq:esn_state_nooutputfeedback}). 
Notably, the maximal local Lyapunov exponent (MLLE) $\lambda$, used to approximate the separation rate in phase space of trajectories having very similar initial states \cite{verstraeten2009quantification}, can be computed from such a matrix.
In autonomous systems, $\lambda<0$ indicates that the system (here ESN) is stable; $\lambda>0$ denotes chaoticity.
A transition point between those two different behaviors is obtained when $\lambda=0$. 
The sign of $\lambda$ provides thus a criterion for detecting the onset of criticality in reservoirs.
Such a criterion is widely used also as a baseline to develop and compare novel criteria \cite{boedecker2012information}.

If reservoir neurons contain a hyperbolic tangent activation, the Jacobian at time $k$ can be conveniently expressed as
\begin{align}
\label{eq:jacob}
&\mathbf{J}(h[k]) = \\
&\nonumber\left[\begin{array}{cccc}
1 - (h_1[k])^2 & 0 & \ldots & 0 \\
0 & 1 - (h_2[k])^2 & \ldots & 0 \\
\vdots & \vdots  & \ddots & \vdots \\
0 & 0 & \ldots & 1 - (h_{N_r}[k])^2 \\
\end{array}\right]
\mathbf{W}_{r}^{r} \ ,
\end{align}
where $h_l[k], l=1, 2, ..., N_r$, is the activation of the $l$-th reservoir neuron at time $k$.
$\lambda$ is then computed by means of the geometric average:
\begin{equation}
\label{eq:MLLE}
\lambda = \max_{n=1, ..., N_r} \frac{1}{K} \sum_{k=1}^{K} \log \left( r_n[k] \right),
\end{equation}
where $r_n[k]$ is the absolute value of the $n$-th eigenvalue of $\mathbf{J}(h[k])$ and $K$ is the total number of samples in the time series under consideration.

Another indicator used to predict the network performance is the minimal singular value of the Jacobian matrix (shortened as mSVJ and denoted in the following as $\eta$), which provides accurate information regarding the ESN dynamics.
The set of hyperparameter configurations that maximize mSVJ gives rise to dynamical systems with good excitability, separating well the input signals in state space \cite{verstraeten2009quantification}.

In this paper, as a means of numerical comparison with the proposed method based on FIM, we will use also MLLE and mSVJ criteria for detecting the onset of criticality in ESNs.

\section{Fisher information matrix and the non-parametric estimator}
\label{sec:FIM_estimator}

The Fisher information matrix \cite{zegers2015fisher} is a symmetric positive semidefinite (PD) matrix whose elements are defined as follows:
\begin{equation}
\label{eq:FIM}
F_{ij}(p_{\boldsymbol{\theta}}(\cdot)) = \int_{\mathcal{D}} p_{\boldsymbol{\theta}}(\mathbf{u}) \left( \frac{\partial\ln p_{\boldsymbol{\theta}}(\mathbf{u})}{\partial \theta_i} \right)\left( \frac{\partial\ln p_{\boldsymbol{\theta}}(\mathbf{u})}{\partial \theta_j} \right) d\mathbf{u},
\end{equation}
where $p_{\boldsymbol{\theta}}(\cdot)$ is a parametric probability density function (PDF), which depends on $d$ parameters $\boldsymbol{\theta}=[\theta_1, \theta_2, ..., \theta_{d}]^{T}\in\Theta\subseteq\mathbb{R}^{d}$ being $\Theta$ the parameter space.
As will be formally discussed in the following sections, in ESN framework $\boldsymbol{\theta}$ contains the hyperparameters under consideration.
In (\ref{eq:FIM}), $\ln p_{\boldsymbol{\theta}}(\cdot)$ represents the log-likelihood function.
For sake of simplicity, we denote $\mathbf{F}(p_{\boldsymbol{\theta}}(\cdot))$ as $\mathbf{F}(\boldsymbol{\theta})$.
The FIM contains $d(d+1)/2$ distinct entries encoding the sensitivity of the PDF with respect to the parameters in $\boldsymbol{\theta}$.

\begin{figure*}[!htp]
\centering
	\subfigure[Thermodynamic systems]{
	\includegraphics[width=0.6\columnwidth, trim={9.5em 2em 13em 1em},clip]{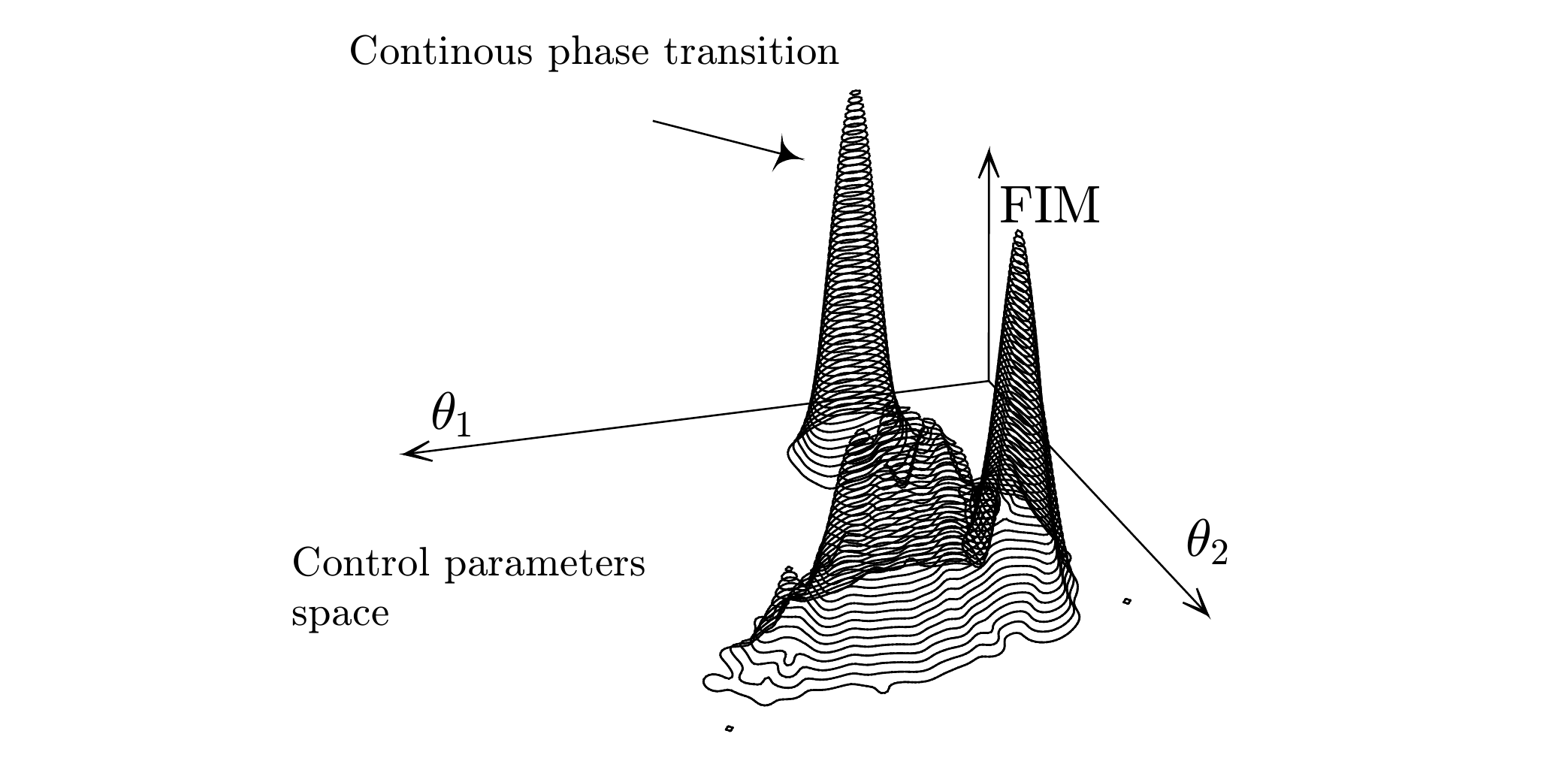}
	}
	~
	\subfigure[Echo State Networks]{
	\includegraphics[width=0.6\columnwidth, trim={9.5em 2em 13em 1em},clip]{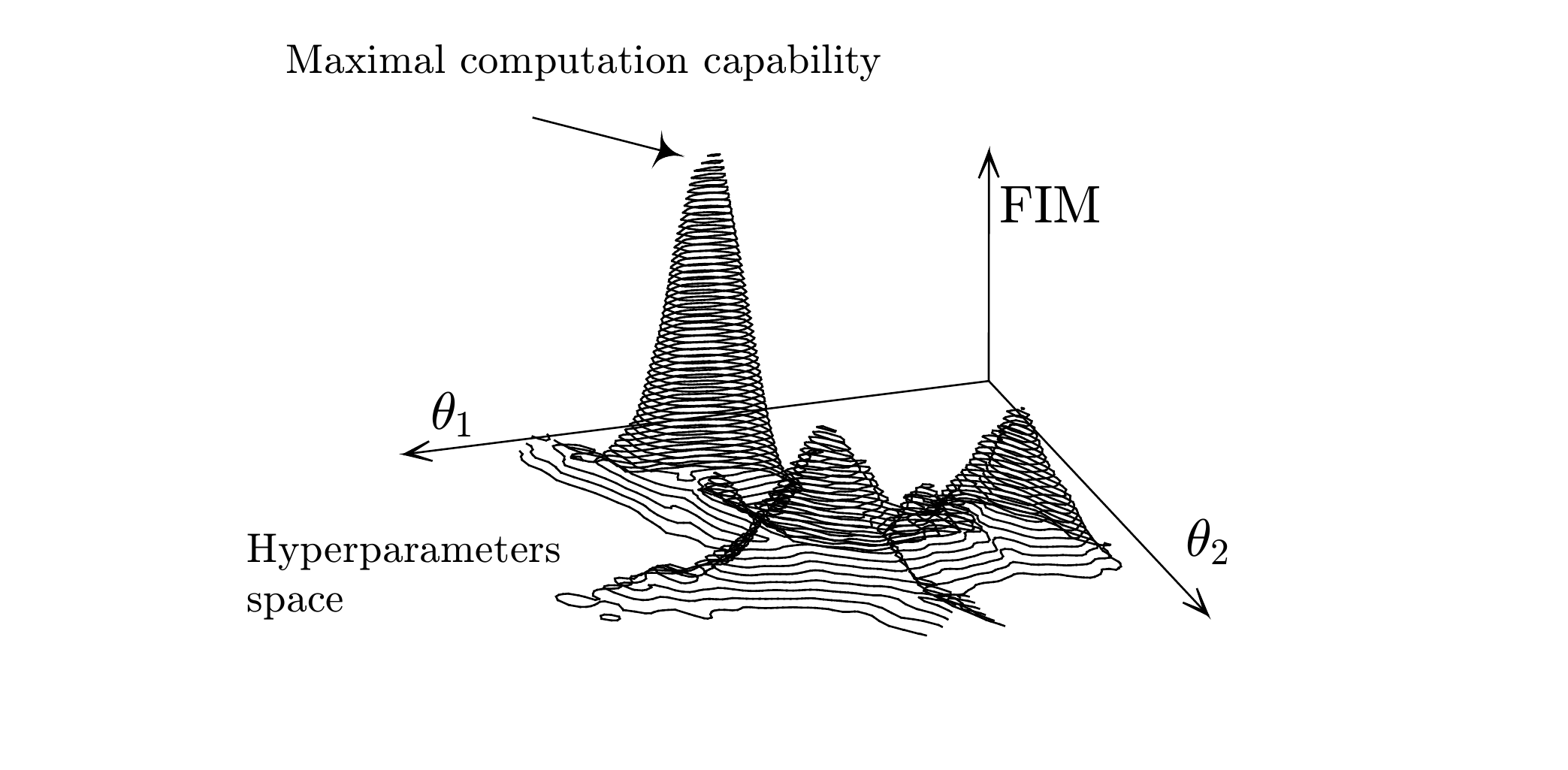}
	}
\caption{The approach based on FIM maximization used to identify a continuous phase transition can be adopted also to characterize dynamics in ESNs. In this context, ESN hyperparameters (e.g., spectral radius, input scaling) play the same role of the control parameters in a thermodynamic system (e.g., temperature affects the magnetization phenomenon). FIM can be used to identify the critical region (edge of criticality) in the ESN hyperparameter space, where the computational capability is maximized.
Note that the densities plotted in the two figures are not related and show the role played by FIM in the two different fields.}
\label{fig:tk}
\end{figure*}

Elements of the FIM can be directly connected with the rate of change of the order parameters of a controlled (thermodynamic) system \cite{prokopenko2011relating}.
An order parameter is a quantity used to discern the phases of a thermodynamic system.
For instance, in the liquid--vapor (first-order) transition of water, temperature acts as a control parameter (at constant pressure), while the difference in density of the two phases -- liquid and gaseous states -- is the order parameter. At the critical temperature, liquid water turns into vapor and the order parameter varies discontinuously.
The mathematical relationship between Fisher information and order parameters is particularly interesting to provide a statistical description of continuous, second-order phase transitions, and, as a consequence, of any complex system approaching a critical transition.
In fact, during a continuous phase transition the order parameter changes continuously. Therefore, differently from first-order transitions, a system can reside and operate in such a critical state. A well-known example of continuous phase transition is the ferromagnetic--paramagnetic transition of iron, where magnetization (the order parameter) is non-zero for temperatures lower than the critical (Curie) one and zero otherwise.
However, second-order derivatives of the observed thermodynamic variable (or, equivalently for continuous transitions, the first-order derivatives of the order parameter) are discontinuous and divergent in at least one dimension. This implies that Fisher information diverges at criticality for infinite systems, while it is maximized in the finite-size system case \cite{prokopenko2011relating}.
This fact provides a clear mathematical justification explaining why the FIM (\ref{eq:FIM}) can be used to detect criticality in complex systems in terms of maximum sensitivity with respect to control parameter changes.
Therefore, as we already mentioned, the critical region (edge of criticality) is a region in parameter space where the Fisher information is maximized; hence we assume here to deal with finite-size systems.
Fig. \ref{fig:tk} provides an intuitive illustration linking criticality and ESNs.

Computation of FIM (\ref{eq:FIM}) requires analytical availability of the PDF.
However, in many experimental settings either (i) the PDF underlying the observed data is unknown or (ii) the relation linking the variation of control parameters $\boldsymbol{\theta}$ on $p_{\boldsymbol{\theta}}(\cdot)$ is unknown.
In a recent paper \cite{6975144}, a non-parametric estimator of the FIM was proposed, which is based on divergence measure
\begin{align}
\label{eq:alpha_divergence}
&D_{\alpha}(p, q) = \\
\nonumber &\frac{1}{4\alpha(1-\alpha)} \int_{\mathcal{D}} \frac{(\alpha p(\mathbf{u}) (1-\alpha)q(\mathbf{u}))^2}{\alpha p(\mathbf{u}) (1-\alpha)q(\mathbf{u})} d\mathbf{u} - (2\alpha -1)^2,
\end{align}
belonging to the family of $f$-divergences; $\alpha\in(0, 1)$; $p(\cdot)$ and $q(\cdot)$ are PDFs both supported on $\mathcal{D}$.

It is well-known \cite{6975144,hidalgo2014information,hidalgo2015cooperation} that the FIM can be approximated by using a proper $f$-divergence measure computed between the parametric PDF of interest and a perturbed version of it. Notably, by expanding with Taylor (\ref{eq:alpha_divergence}) up to the second order we obtain:
\begin{equation}
\label{eq:divergence_FIM}
D_{\alpha}(p_{\boldsymbol{\theta}}, p_{\hat{\boldsymbol{\theta}}}) \simeq \frac{1}{2}\mathbf{r}^{T} \mathbf{F}(\boldsymbol{\theta}) \mathbf{r},
\end{equation}
where $\hat{\boldsymbol{\theta}}=\boldsymbol{\theta}+\mathbf{r}$, being $\mathbf{r} \sim \mathcal{N}(\mathbf{0}, \sigma^2\mathbf{I}_{d\times d})$ a small normally distributed perturbation vector with standard deviation $\sigma$.

Divergence (\ref{eq:alpha_divergence}) can be computed directly without the need to estimate the PDFs by means of an extension of the Friedman-Rafsky multi-variate two-sample test statistic \cite{friedman1979multivariate}.
The test operates by using two datasets, $\mathcal{S}_{p}$ and $\mathcal{S}_{q}$, each one containing samples extracted from $p(\cdot)$ and $q(\cdot)$, respectively.
Theorem 1 in \cite{6975144} shows that, as the number of samples $n=|\mathcal{S}_{p}|$ and $m=|\mathcal{S}_{q}|$ increases asymptotically, we have:
\begin{equation}
\label{eq:FR_test}
1-\mathcal{C}(\mathcal{S}_{p}, \mathcal{S}_{q})\frac{n+m}{2nm}\ \xrightarrow{a.s.}\ D_{\alpha}(p, q),
\end{equation}
where $\mathcal{C}(\mathcal{S}_{p}, \mathcal{S}_{q})$ is the outcome (expected to be normally distributed) of the Friedman-Rafsky test, which basically provides a way to measure the similarity between two datasets.
Interestingly, such a test allows to analyze also high-dimensional data, since it makes use of a graph-based representation of the data (a minimum spanning tree).

In the following, for the sake of brevity we omit $\boldsymbol{\theta}$ in most of the equations and refer to the estimated FIM as $\mathbf{\hat{F}}$.
\cite{6975144} proposes two different approaches for estimating the FIM (\ref{eq:divergence_FIM}).
The first one is based on the well-known least-square optimization:
\begin{equation}
\label{eq:LS_opt}
\mathbf{\hat{F}}_{\mathrm{hvec}} = (\mathbf{R}^{T}\mathbf{R})^{-1} \mathbf{R}^{T} \mathbf{v}_{\boldsymbol{\theta}},
\end{equation}
where $\mathbf{v}_{\boldsymbol{\theta}}=[v_{\boldsymbol{\theta}}(\mathbf{r}_1), ..., v_{\boldsymbol{\theta}}(\mathbf{r}_M)]^T$, with $v_{\boldsymbol{\theta}}(\mathbf{r}_i)=2D_{\alpha}(p_{\boldsymbol{\theta}}, p_{\hat{\boldsymbol{\theta}_i}}), i=1, ..., M$, and $D_{\alpha}(\cdot, \cdot)$ is computed by means of the left-hand side of (\ref{eq:FR_test}).
$\mathbf{R}$ is a matrix containing all $M$ perturbation vectors $\mathbf{r}_i$ arranged as column vectors, and $\mathbf{\hat{F}}_{\mathrm{hvec}}$ is the half-vector representation of $\mathbf{\hat{F}}$.
Note that a vector representation $\mathbf{\hat{F}}_{\mathrm{vec}}$ of $\mathbf{\hat{F}}$ reads as $\left[f_{11}, \ldots, f_{m1}, f_{12}, \ldots, f_{mn}\right]^T$.
Since $\mathbf{\hat{F}}$ is symmetric, it can be represented through the half-vector representation, $\mathbf{\hat{F}}_{\mathrm{hvec}}$, which is obtained by eliminating all superdiagonal elements of $\mathbf{\hat{F}}$ from $\mathbf{\hat{F}}_{\mathrm{vec}}$ \cite{magnus1995matrix}.
$\mathbf{\hat{F}}_{\mathrm{hvec}}$ in (\ref{eq:LS_opt}) is hence defined as $\left[\hat{f}_{11}, \ldots, \hat{f}_{dd}, \hat{f}_{12}, \ldots, \hat{f}_{d(d-1)}\right]^T$, where the diagonal elements are located in the first components of the vector.

However, the least-square approach (\ref{eq:LS_opt}) does not guarantee to find an approximation of the FIM which is PD.
A second approach requires solving a semidefinite optimization problem, which assures that the resulting FIM is PD:
\begin{equation}
\label{eq:PSD_opt}
\begin{aligned}
& \underset{\mathbf{F}_{\mathrm{hvec}}}{\text{minimize}}
& & \lVert \mathbf{R} \mathbf{F}_{\mathrm{hvec}} - \mathbf{v}_{\boldsymbol{\theta}} \rVert^{2} \\
& \text{subject to}
& & \mathbf{F}_{\mathrm{hvec}}(i) = \mathrm{diag}\left(\mathrm{mat}\left(\mathbf{\hat{F}}_{\mathrm{hvec}}\right)\right), \ i \in \{1, \ldots, d\}, \\
& & & \mathrm{mat}\left(\mathbf{F}_{\mathrm{hvec}}\right) \succeq \mathbf{0}_{d\times d}.
\end{aligned}
\end{equation}

The $\mathrm{diag}(\cdot)$ operator returns the diagonal elements of a matrix  and the $\mathrm{mat}(\cdot)$ operator converts the argument from a vector form into a square $d\times d$ matrix.
The diagonal values of the FIM as expressed by the first constraint are computed through the least-square optimization (\ref{eq:LS_opt}).
The second constraint, instead, guarantees the estimated matrix to be PD.

Such a convex optimization problem (\ref{eq:PSD_opt}) can be implemented by using the framework provided in \cite{cvx,gb08}.
However, there, a non-trivial implementation in matrix form of the second constraint, i.e., $\mathrm{mat}\left(\mathbf{F}_{\mathrm{hvec}}\right) \succeq \mathbf{0}_{d\times d}$, must be provided to define a proper semidefinite problem.
In this paper, we solve this issue and provide a novel method granting $\mathrm{mat}\left(\mathbf{F}_{\mathrm{hvec}}\right) \succeq \mathbf{0}_{d\times d}$ (see demonstration in Appendix \ref{sec:constraint_implementation}).

\section{Critical region identification for ESNs}
\label{sec:proposed_method}

\begin{figure*}[htp!]
\center
\scalebox{0.7}{%
\begin{tikzpicture}
[
 inner/.style={rounded corners=5pt,draw,fill=blue!5,thick, align=center},
 outer/.style={draw=gray,dashed,fill=blue!5,thick},
 var/.style={draw,fill=gray!30,thick},
 operator/.style={draw,fill=gray!2,thick},
 outer2/.style={draw=black,dashed,fill=none,thick},
]
     
\node [inner, minimum height=4em, minimum width=7.5em] (N1)  at (-10em,0em) {Collect ESN \\ activations \\ $\mathcal{S}_{\boldsymbol{\theta}}=\{\mathbf{h}[k]\}_{k=1}^{K}$};
\node [inner, minimum height=4em, minimum width=7.5em] (N2)  at (0em,0em) {Non-parametric \\  estimation \\ of FIM $\hat{\mathbf{F}}(\boldsymbol{\theta})$};
\node [inner, minimum height=4em, minimum width=7.5em] (N3)  at (10em,0em) {Evaluate \\ determinant \\ of $\hat{\mathbf{F}}(\boldsymbol{\theta})$};
\node [outer2, minimum height=10em, minimum width=30em] (N4)  at (0em,-1.5em) {};

\node[var, align=center, minimum width=8.1em] (N5)  at  (-20.5em,0.5em) {Initial parameter \\ configuration $\boldsymbol{\theta}_0$};
\node[var, align=center, minimum width=8.1em] (N6)  at  (-20.5em,-3.5em) {Input signal \\ $\mathbf{x}[1], \ldots,\mathbf{x}[K]$};
\node[rounded corners=8pt, var, align=center] (N7)  at  (22em,-1em) {$\argmax \limits_{\boldsymbol{\theta}^* \in \Theta} \; \mathrm{det}(\hat{\mathbf{F}}(\boldsymbol{\theta}^*))$};

\node[outer sep=0,inner sep=0,minimum size=0] (n0)  at  (0em,-4.3em) {Select new ESN hyperparameters $\boldsymbol{\theta}$};
\node[outer sep=0,inner sep=0,minimum size=0] (n1)  at  (10em,-5em) {};
\node[outer sep=0,inner sep=0,minimum size=0] (n2)  at  (-10em,-5em) {};
\node[outer sep=0,inner sep=0,minimum size=0] (n3)  at  (-15em,0.5em) {};
\node[outer sep=0,inner sep=0,minimum size=0] (n4)  at  (-15em,-3.5em) {};
\node[outer sep=0,inner sep=0,minimum size=0] (n5)  at  (15em,-1em) {};

\draw  (N3) edge (n1);
\draw  (n1) edge (n2);
\draw  [-latex](n2) edge (N1);
\draw  [-latex](N1) edge (N2);
\draw  [-latex](N2) edge (N3);
\draw  [-latex](N5) edge (n3);
\draw  [-latex](N6) edge (n4);
\draw  [-latex](n5) edge (N7);

\end{tikzpicture}
}
\caption{Schematic, high-level description of the proposed procedure.}
\label{fig:algomaxFIMdet}
\end{figure*}
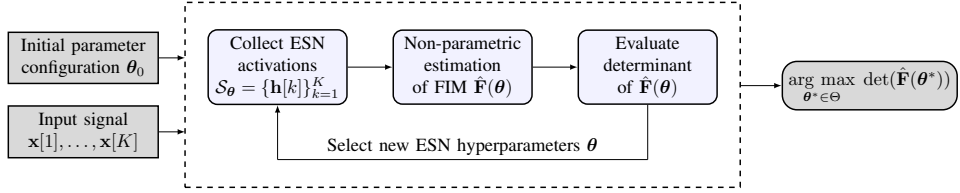

Our goal is to find the edge of criticality, i.e., a region in parameter space $\mathcal{K}\subset\Theta$ where the ESN computational capability is maximized.
Fig. \ref{fig:algomaxFIMdet} shows a schematic description of the main phases involved in the proposed method.
%

Let us discuss in details the proposed procedure.
In order to determine $\mathcal{K}$, we propose an algorithm exploiting the FIM properties of a system undergoing a continuous phase transition.
FIM defines a metric tensor for the smooth manifold of parametric PDFs embedded in $\Theta$ \cite{prokopenko2011relating}, allowing thus also for a geometric characterization of the system under analysis.
It is possible to prove \cite{mastromatteo2011criticality} that $\mathcal{K}$ corresponds to a region in $\Theta$ characterized by the largest volume (high concentration of parametric PDF). In addition, it worth mentioning that, if $\Theta\subseteq\mathbb{R}^d$, then the related edge of criticality $\mathcal{K}$ is a $d-1$ manifold embedded in $\Theta$.
This geometric result can be exploited by using the determinant, $\mathrm{det}(\mathbf{F}(\boldsymbol{\theta}))$, which is monotonically related to the aforementioned volume in the parameter space.
Therefore, considering that the FIM is a PD matrix, and hence its determinant is always non-negative, we identify $\mathcal{K}$ with all those hyperparameters $\boldsymbol{\theta}^*$ for which:
\begin{equation}
\label{eq:config_maximum_FIMdet}
\boldsymbol{\theta}^*=\argmax_{\boldsymbol{\theta}\in\Theta} \mathrm{det}(\mathbf{F}(\boldsymbol{\theta})).
\end{equation}

Algorithm \ref{alg:eoc_single_param} delivers the pseudo-code of the proposed procedure.
As said before, the impact provided by the variation of the control parameters $\boldsymbol{\theta}$ on the resulting ESN state cannot be described analytically without making further assumptions \cite{massar2013mean}. In fact, the (unknown) input signal driving the network plays an important role in the resulting ESN dynamics.
Therefore, in order to calculate $\mathbf{F}(\boldsymbol{\theta})$, in Algorithm \ref{alg:eoc_single_param} we rely on the non-parametric FIM estimator described in Sec. \ref{sec:FIM_estimator}.
The estimation of the FIM for a given $\boldsymbol{\theta}$ is performed by analyzing the sequence $\mathcal{S}_{\boldsymbol{\theta}}=\{\mathbf{h}[k]\}_{k=1}^{K}$ of reservoir neuron activations produced during the processing of a given input $\mathbf{x}$ of length $K$.
Since $\mathbf{h}[k]\in[-1, 1]^{N_r}$, the domain of the PDF in (\ref{eq:FIM}) is defined as $\mathcal{D}=[-1, 1]^{N_r}$.
Additional sequences of activations, $\mathcal{S}_{\hat{\boldsymbol{\theta}}_j}$, are considered (see line \ref{alg:perturbations}), which are obtained by perturbing $M$ times the current network configuration $\boldsymbol{\theta}$ under analysis, and processing the same input $\mathbf{x}$.
Perturbations are modeled with a zero-mean noise with a spherical covariance matrix, thus characterized by a single scalar parameter $\sigma$ controlling the magnitude of the perturbation.
In this paper, we estimate the FIM by solving the optimization problem (\ref{eq:PSD_opt}) according to our formulation as described in Appendix \ref{sec:constraint_implementation}.
In order to make the estimation more robust, we follow an ensemble approach and perform a number of independent trials (see line \ref{alg:trials}). The determinant is computed only once on the resulting average FIM, which is obtained by using $T$ independent random realizations of the ESN architecture chosen for the experiment (see line \ref{alg:average_FIM}).

In theory, the parameter space $\Theta$ is continuous.
However, here we assume that the parameter space $\Theta$ is quantized according to some user-defined resolution.
Although this is not a necessary assumption for the proposed methodology, it allows us to disentangle the problems of defining from finding the edge of criticality. In fact, our main goal here is to provide a principled definition of the critical region characterizing the ESN (hyper-)parameter space and related behaviors. More efficient and/or accurate search schemes will be considered in future research studies.
Accordingly, the criterion in (\ref{eq:config_maximum_FIMdet}) identifies a ``quantized'' critical region $\mathcal{K}$ in $\Theta$ represented by a single hyperparameter configuration, $\boldsymbol{\theta}^*$.
\begin{algorithm}[h!]\footnotesize
\caption{Procedure for determining an ESN configuration on the edge of criticality.}
\label{alg:eoc_single_param}
\begin{algorithmic}[1]
\REQUIRE An ESN architecture, input $\mathbf{x}$ of $K$ samples, quantized parameter space $\Theta$, standard deviation $\sigma$ for the perturbations, number of trials $T$ and perturbations $M$.
\ENSURE A configuration $\boldsymbol{\theta}^*\in\mathcal{K}$ 
\STATE Select an initial parameter configuration, $\boldsymbol{\theta}\in\Theta$; maximum $\eta=0$
\LOOP
\FOR{$t=1$ to $T$}\label{alg:trials}
\STATE Randomly initialize the ESN weight matrices
\STATE Configure ESN with $\boldsymbol{\theta}$ and process input $\mathbf{x}$
\STATE Collect the related activations $\mathcal{S}_{\boldsymbol{\theta}}=\{\mathbf{h}[i]\}_{i=1}^{K}$
\FOR{$j=1$ to $M$}\label{alg:perturbations}
\STATE Generate a perturbation vector $\mathbf{r}_j \sim \mathcal{N}(\mathbf{0}, \sigma^2\mathbf{I}_{d\times d})$\label{alg:perturbation_vector}
\STATE Randomly initialize the ESN weight matrices
\STATE Configure ESN with perturbed version $\hat{\boldsymbol{\theta}}_j=\boldsymbol{\theta}+\mathbf{r}_j$ and process input $\mathbf{x}$
\STATE Collect the related activations $\mathcal{S}_{\hat{\boldsymbol{\theta}}_j}=\{\mathbf{h}[i]\}_{i=1}^{K}$
\ENDFOR
\STATE Define $\mathcal{S}_{\hat{\boldsymbol{\theta}}}=\cup_{j=1}^{M}\mathcal{S}_{\hat{\boldsymbol{\theta}}_j}$
\STATE Estimate the FIM $\mathbf{F}^{(t)}(\boldsymbol{\theta})$ of trial $t$ using $\mathcal{S}_{\boldsymbol{\theta}}$ and $\mathcal{S}_{\hat{\boldsymbol{\theta}}}$ with the non-parametric estimator introduced in Sec. \ref{sec:FIM_estimator}
\ENDFOR
\STATE Compute the average FIM, $\mathbf{F}(\boldsymbol{\theta})$, using all $\mathbf{F}^{(t)}(\boldsymbol{\theta}), t=1, ..., T$\label{alg:average_FIM}
\IF{$\mathrm{det}(\mathbf{F}(\boldsymbol{\theta}))>\eta$}
\STATE Update $\eta=\mathrm{det}(\mathbf{F}(\boldsymbol{\theta}))$ and $\boldsymbol{\theta}^*=\boldsymbol{\theta}$
\ENDIF
\IF{Stop criterion is met}\label{algo:stop_criterion}
\RETURN $\boldsymbol{\theta}^*$
\ELSE
\STATE Select a new $\boldsymbol{\theta}\in\Theta$ based on a suitable search scheme\label{algo:search_criterion}
\ENDIF
\ENDLOOP
\end{algorithmic}
\end{algorithm}

\subsection{Analysis of computational complexity}
\label{sec:comp_complexity}

The asymptotic computational complexity (including also constant terms) of Algorithm \ref{alg:eoc_single_param} can be summarized as follows (assuming a grid search):
\begin{equation}
\label{eq:comp_complexity}
O(G (T (N_r^2 + KN_r + M (N_r^2 + KN_r) + E_{\mathrm{FIM}} ) + d^3 + Td^2 ) ).
\end{equation}

In \ref{eq:comp_complexity}, $G$ is the number of hyperparameter configurations taken into account, $T$ is the number of trials, $N_r$ is the number of neurons in the reservoir, $K$ is the input signal length, and $M$ is the number of perturbations.
The cost related to the computation of the determinant of FIM is hence $O(d^3)$, where $d$ is the number of hyperparameters taken into account.
The last term, $Td^2$, accounts for the computation of the average FIM.
The $E_{\mathrm{FIM}}$ term describes the complexity of the non-parametric FIM estimator described in Sec. \ref{sec:FIM_estimator}.
$E_{\mathrm{FIM}}$ cost can be decomposed in two different terms.
First, \ref{eq:FR_test}, the computation of $\alpha$ divergence, has a cost that is given by the MST computation on $z=(M+1)K$ samples, that is bounded by $O(z^2\log(z))$.
That cost is multiplied by $M$, the number of perturbations.
Second, the cost associated with solving the optimization problem shown in \ref{eq:PSD_opt2}.
The computational complexity of the constraint satisfaction is bounded by $d^2$.
The semidefinite optimization program can be solved in polynomial time, i.e., $O(d^p)$, where $p$ is some positive integer \cite{cvx,gb08}.

Typically, $d$ is much smaller than both $N_r$ and $K$. Therefore, polynomial terms in $d$ do not pose a problem from the computational complexity viewpoint.
The computation complexity (\ref{eq:comp_complexity}) is hence dominated by the $E_{\mathrm{FIM}}$ cost.

\section{Experiments}
\label{sec:exps}

In this section, we evaluate the effectiveness of the proposed method based on FIM for determining ESN hyperparameter configurations lying on the edge of criticality.
The proposed method is firstly validated on a set of benchmarks used in the ESN literature.
In particular, we consider the short-term memory capacity (Sec. \ref{sec:MC_test}) and then a forecast task on different time series models (Sec. \ref{sec:prediction_test}).
For such benchmarks, the training set consists of 5000 samples, while 500 samples are used for testing.
Successively, in Sec. \ref{sec:orange_call_load} we validate the proposed methodology on a real-world application involving the prediction of time series related to phone calls load \cite{bianchi2015prediction}. Here, the training set consists of 3335 samples, while 500 samples are used for testing.

The hyperparameters are selected in a discretized space through a grid search, which considers 10 different configurations for each parameter.
Specifically, we search for the spectral radius $\theta_{\mathrm{SR}}$ in $[0.4, 1.6]$, input scaling $\theta_{\mathrm{IS}}$ in $[0.3, 0.8]$, and reservoir connectivity $\theta_{\mathrm{RC}}$ in $[0.1, 0.7]$, evaluating a total of 1000 hyperparameter configurations. Such intervals have been chosen by focusing on the ranges that produce relevant variations in the network behavior.
For each hyperparameter configuration, in Algorithm \ref{alg:eoc_single_param} we perform $T=10$ independent trials and $M=80$ perturbations to compute the ensemble average of the FIM; the variance for the perturbations is set to $\sigma^2=0.25$.
In each trial, we sample new (and independent) input and reservoir connection weights ($\mathrm{W}_i^r$ and $\mathrm{W}_r^r$).
The readout layer is trained by using a standard ridge least-square regression, with the regularization parameter set to $0.05$.
For each test we use a reservoir with $N_r = 100$ neurons; a standard drop-out procedure is adopted \cite{jaeger2002adaptive}, which discards the first 100 states not to consider ESN transient.

\begin{figure*}[!htp]
\centering

	\subfigure[MC test]{
	\includegraphics[width=0.45\columnwidth, trim={0em 0em 0em 0em},clip]{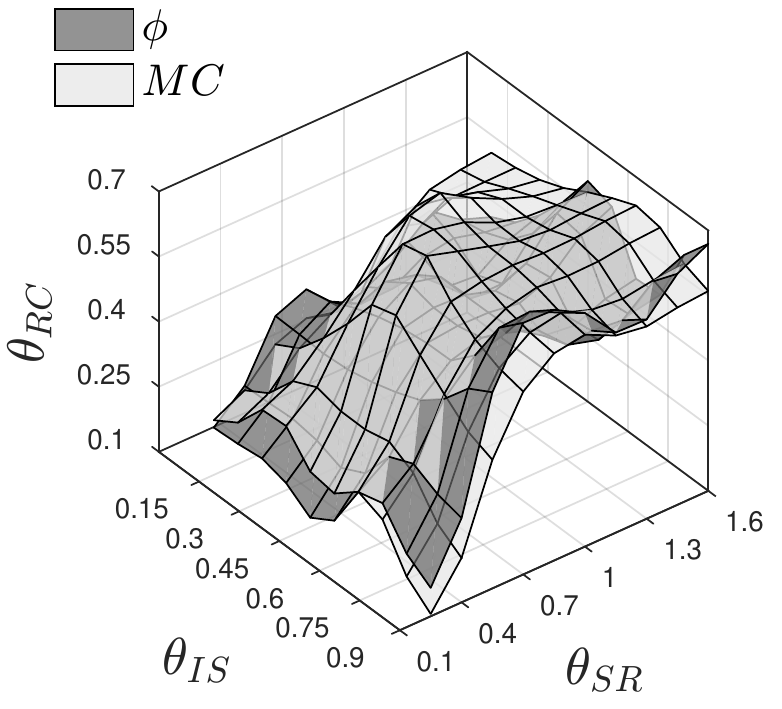}
	\includegraphics[width=0.45\columnwidth, trim={0em 0em 0em 0em},clip]{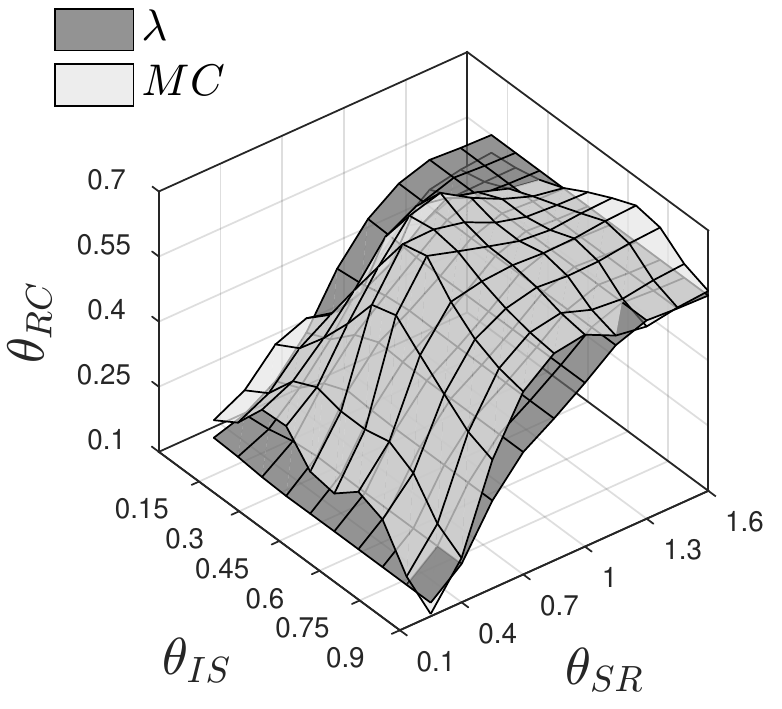}
	\includegraphics[width=0.45\columnwidth, trim={0em 0em 0em 0em},clip]{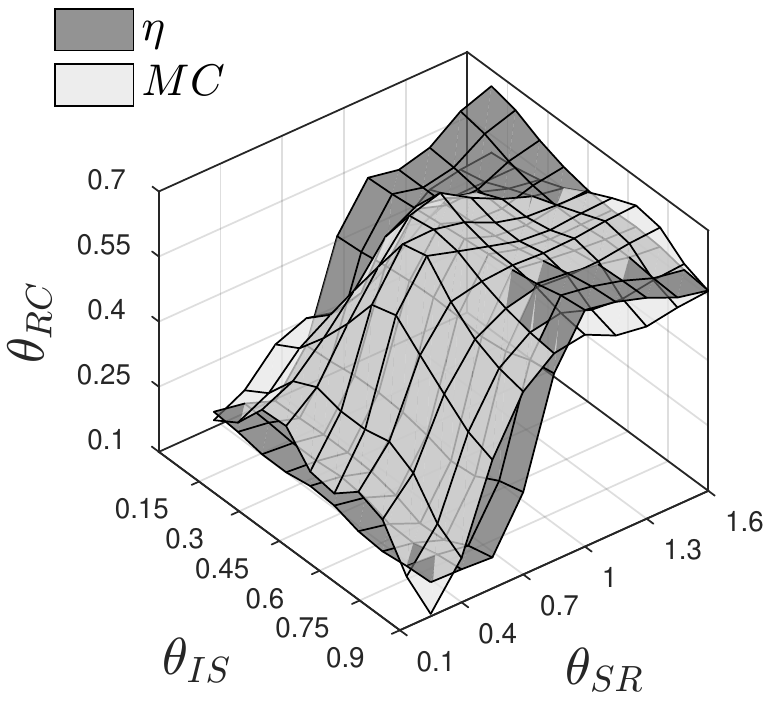}
	\label{fig:MC_task}
	}
	
	\subfigure[SIN prediction task]{
	\includegraphics[width=0.45\columnwidth, trim={0em 0em 0em 0em},clip]{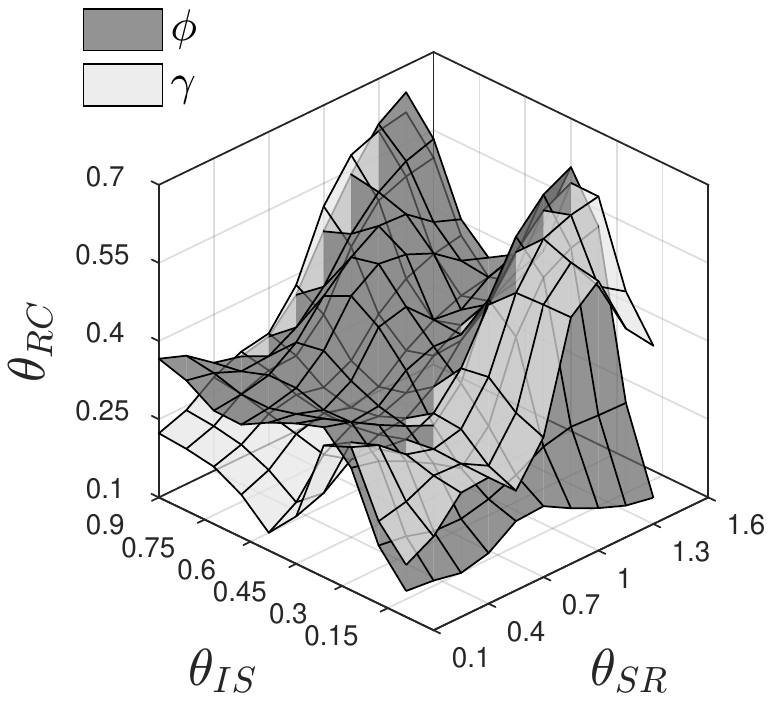}
	\includegraphics[width=0.45\columnwidth, trim={0em 0em 0em 0em},clip]{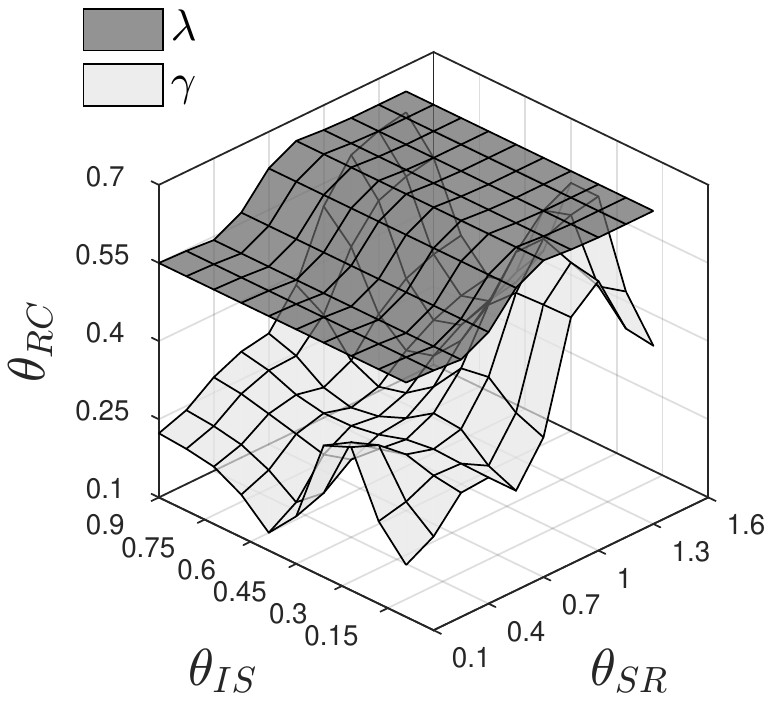}
	\includegraphics[width=0.45\columnwidth, trim={0em 0em 0em 0em},clip]{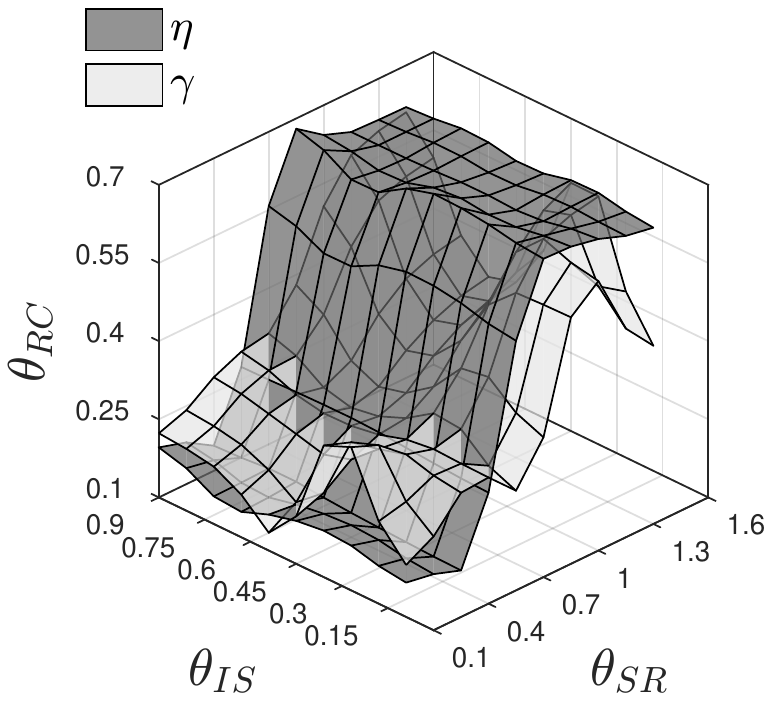}
	\label{fig:SIN_task}
	}
	
	\subfigure[MG prediction task]{
	\includegraphics[width=0.45\columnwidth, trim={0em 0em 0em 0em},clip]{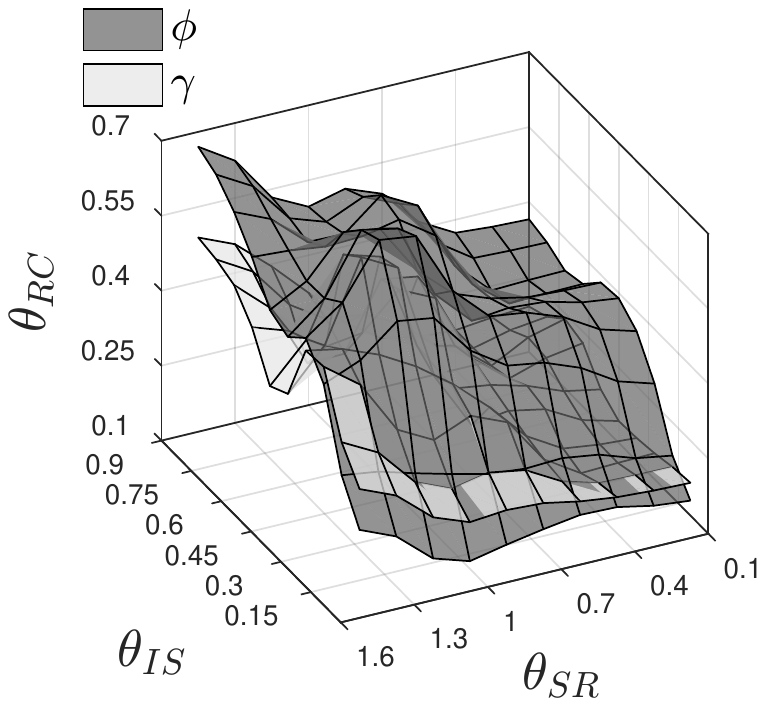}
	\includegraphics[width=0.45\columnwidth, trim={0em 0em 0em 0em},clip]{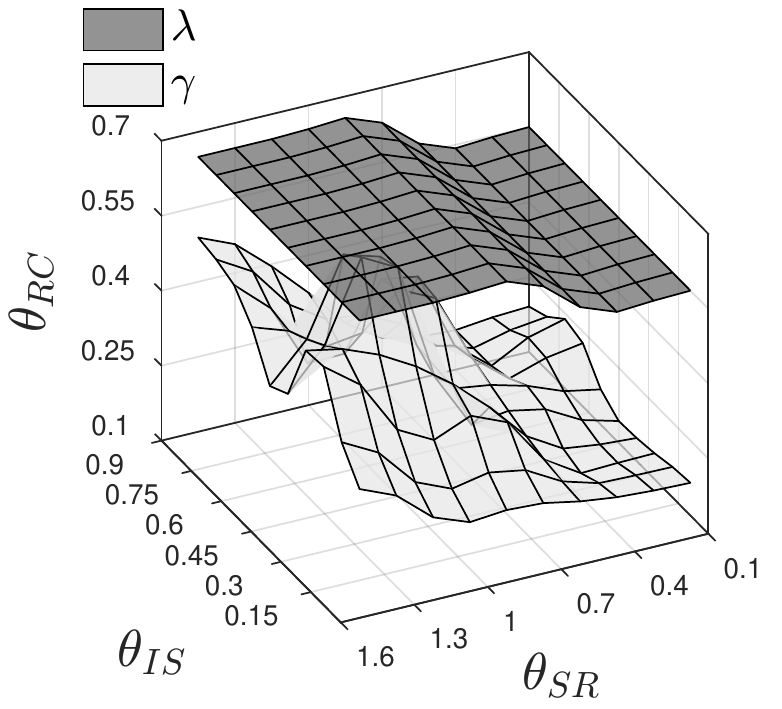}
	\includegraphics[width=0.45\columnwidth, trim={0em 0em 0em 0em},clip]{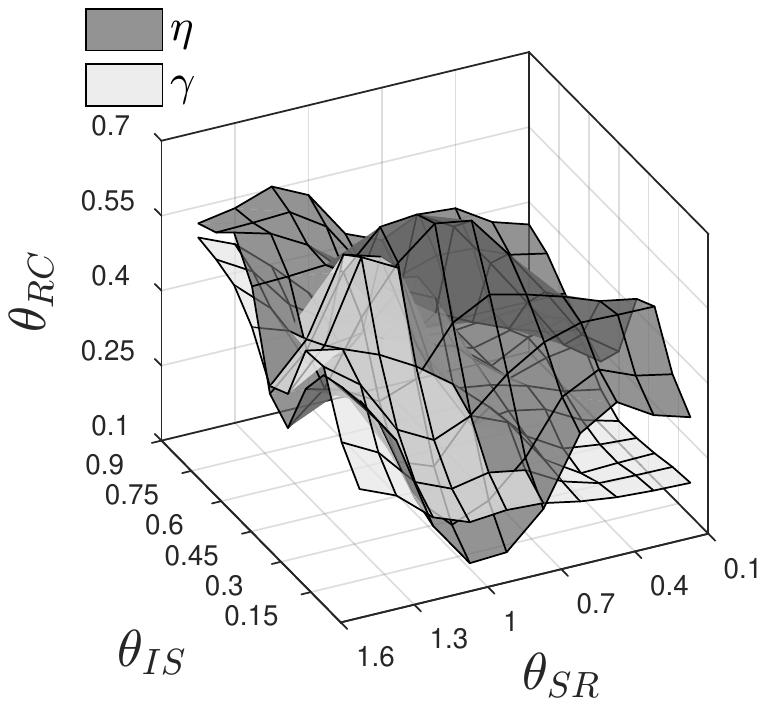}
	\label{fig:MG_task}
	}
	
	\subfigure[NARMA prediction task]{
	\includegraphics[width=0.45\columnwidth, trim={0em 0em 0em 0em},clip]{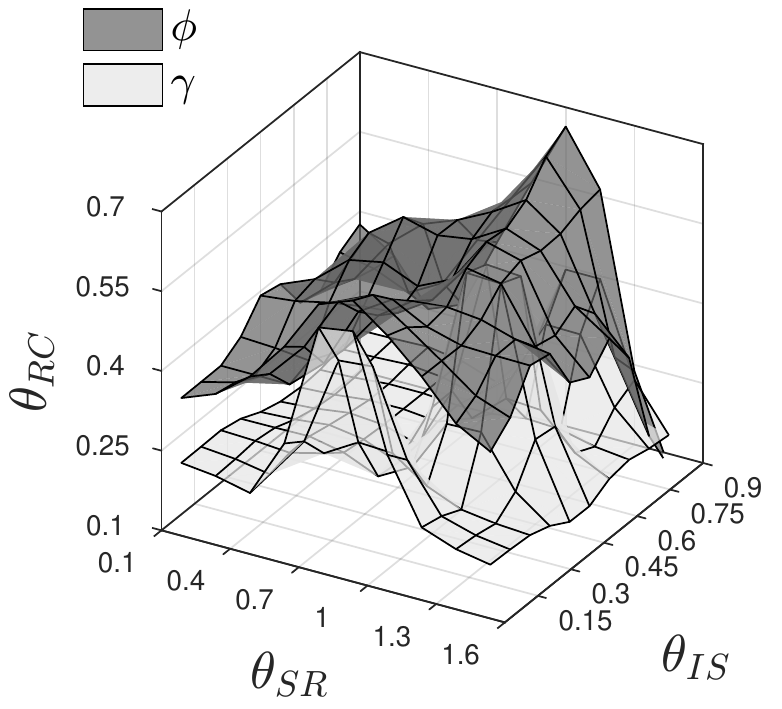}
	\includegraphics[width=0.45\columnwidth, trim={0em 0em 0em 0em},clip]{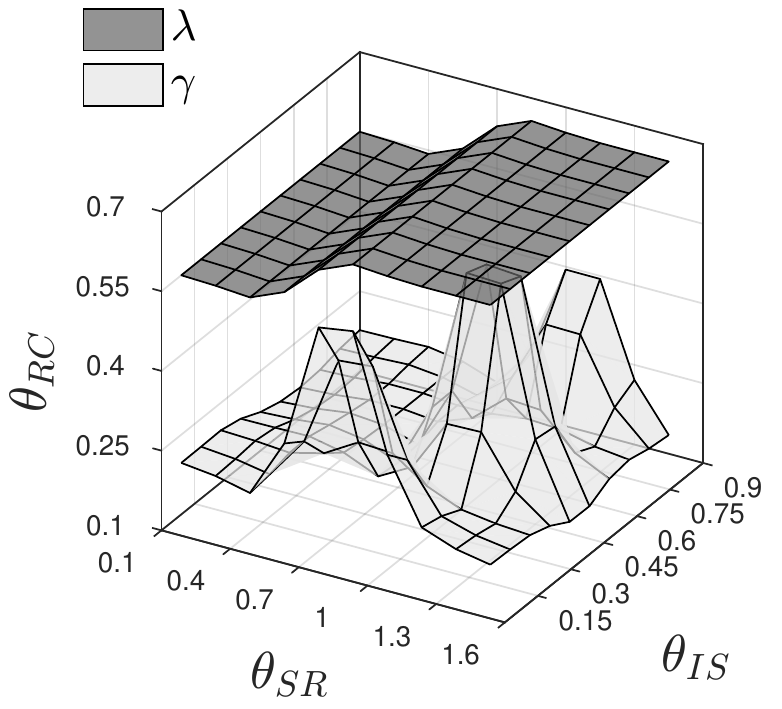}
	\includegraphics[width=0.45\columnwidth, trim={0em 0em 0em 0em},clip]{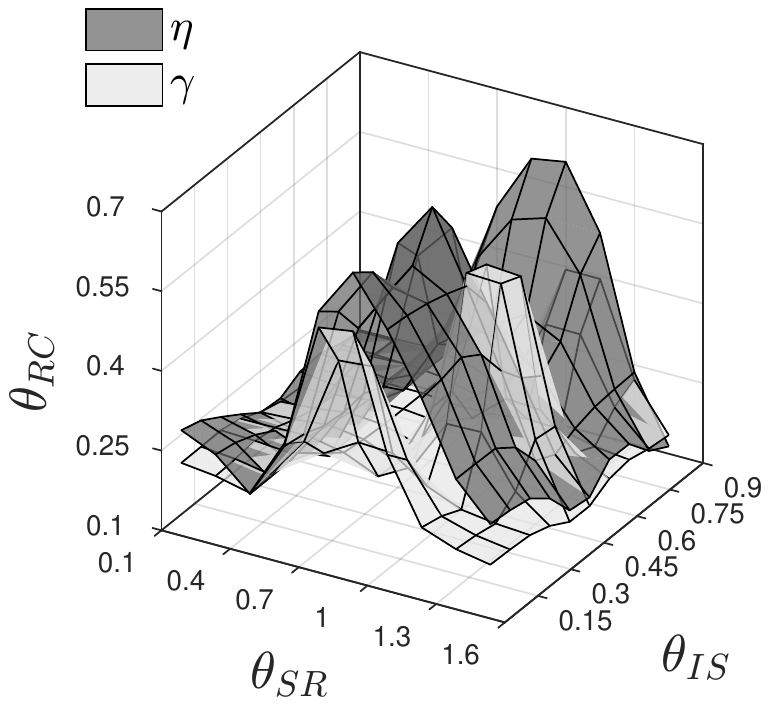}
	\label{fig:NARMA_task}
	}
	
	\subfigure[D4D prediction task]{
	\includegraphics[width=0.45\columnwidth, trim={0em 0em 0em 0em},clip]{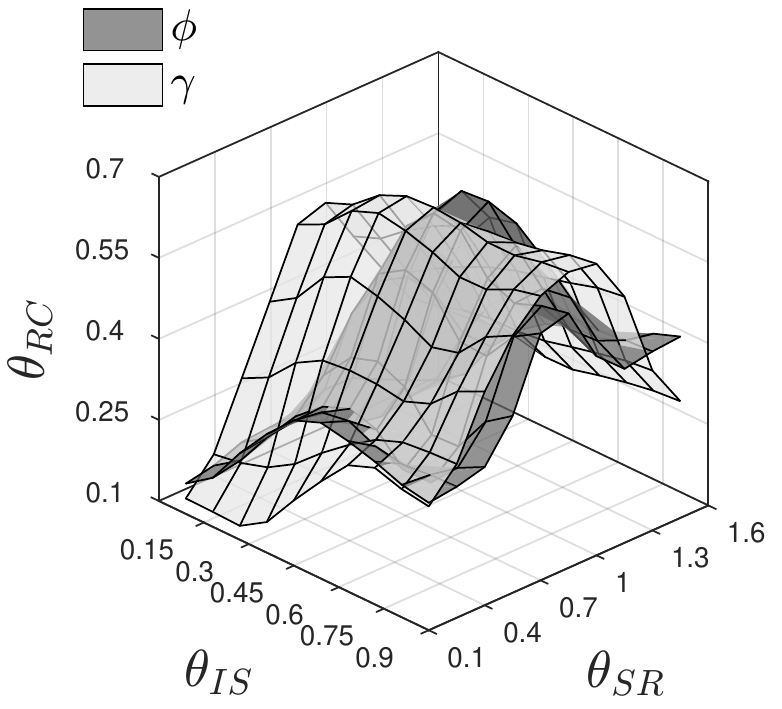}
	\includegraphics[width=0.45\columnwidth, trim={0em 0em 0em 0em},clip]{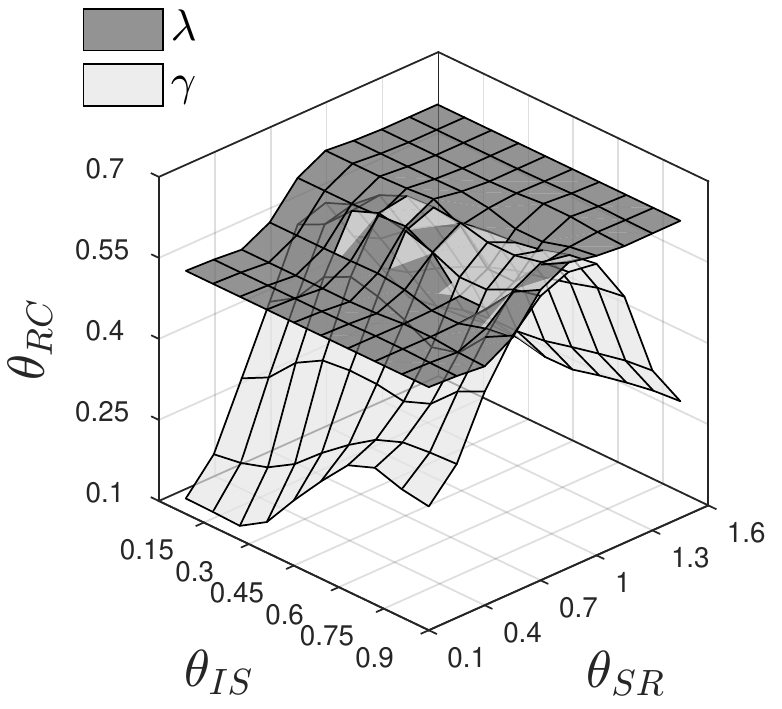}
	\includegraphics[width=0.45\columnwidth, trim={0em 0em 0em 0em},clip]{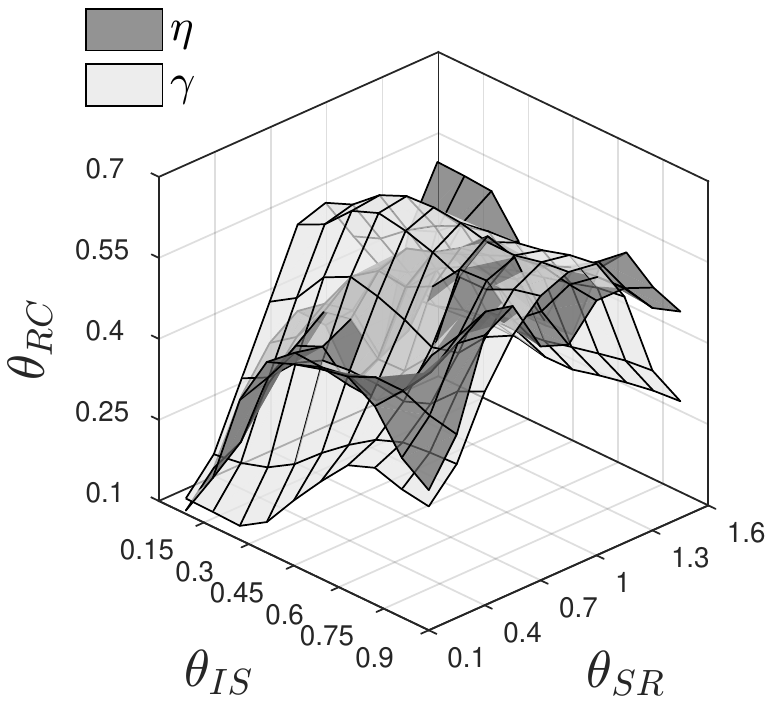}
	\label{fig:ORANGE_task}
	}

\caption{
In each figure, we graphically represent the computed edge of criticality for each of the considered methods. The light gray manifold represents configurations of spectral radius ($\theta_{SR}$), input scaling ($\theta_{IS}$), and reservoir connectivity ($\theta_{RC}$) that maximize Memory Capacity (MC) or prediction accuracy ($\gamma$).
The dark gray manifolds represent (from left to right): configurations where the FIM determinant is maximized ($\phi$); configurations where MLLE crosses zero ($\lambda$); configurations where mSVJ is maximized ($\eta$).}
\label{fig:tests}
\end{figure*}

In Fig. \ref{fig:tests}, we report the critical regions of the parameter space identified in each test by maximization of FIM determinant, zero-crossing of MLLE, and maximization of mSVJ. For the sake of brevity, we refer to these regions as $\phi$, $\lambda$, and $\eta$, respectively.
The light gray manifold corresponds to the regions in parameter space where the performance of the network is maximized and the dark gray manifolds represent $\phi$, $\lambda$, and $\eta$.
In Tab. \ref{tab:results}, we report the numerical values of the correlations between the light gray manifold and the dark gray ones.
In the following subsections, we discuss the details of obtained results.

\subsection{Memory capacity}
\label{sec:MC_test}

This test quantifies the capability of ESN to remember previous inputs, relative to an i.i.d. signal.
Given a time delay $\delta>0$, here we train an ESN to reproduce at time $k$ the input $\mathbf{x}[k-\delta]$.
Memory Capacity (MC) is measured as the squared correlation coefficient between the desired output, which is the input signal delayed by different $\delta$ time steps, and the observed network output $\mathbf{y}[k]$:
\begin{equation}
\label{eq:MC}
\mathrm{MC} = \sum \limits_{\delta=1}^{\delta_{\text{max}}} \frac{\text{cov}^2\left( \mathbf{x}[k-\delta], \mathbf{y}[k] \right)}{\mathrm{var}\left(\mathbf{x}[k-\delta]\right) \mathrm{var}\left(\mathbf{y}[k]\right)}.
\end{equation}
MC is computed by training several readout layers, one for each delay $\delta \in \{1, 10, \ldots, 100\}$, while keeping fixed input and reservoir layers. 

As it is possible to notice in Fig. \ref{fig:MC_task}, the critical regions identified by each one of the three methods follow the region of the hyperparameter space where MC is maximized with good accuracy.
The degrees of correlation for the MC task are provided in Tab. \ref{tab:results}. It is interesting to note that $\lambda$ shows a very high correlation (81\%) preforming better than $\eta$ for this task. The correlation between $\phi$ and the region with maximum MC is also very high (75\%), showing that both $\phi$ and $\lambda$ can be used as reliable indicators to identify the optimal configurations that enhance the short-term memory capacity of ESNs. The $p$-values for each correlation measure are lower than $0.05$, indicating statistical significance of the results.

\subsection{Prediction accuracy on benchmarks}
\label{sec:prediction_test}

In this test, we evaluate the effectiveness of using $\phi$, $\lambda$, and $\eta$ to identify regions of hyperparameters where prediction accuracy is maximal.
We define the prediction accuracy as $\gamma=\max\{1-\mathrm{NRMSE}, 0\}$, were NRMSE is the Normalized Root Mean Squared Error of the ESN.
The accuracy is evaluated on three prediction tasks of increasing difficulty.
For each of them, we set the forecast step $\tau_f>0$ equal to the smallest time lag that guarantees input measurements to be decorrelated, which corresponds to the first zero of the autocorrelation function of the input signal.

\textbf{Sinusoidal input.}
In the first test, the ESN is trained to predict a sinusoidal (SIN) input using a forecast step equal to 1/4 of its period.
In Fig. \ref{fig:SIN_task}, both $\phi$ and $\eta$ are consistent with $\gamma$, while $\lambda$ shows a lower agreement.
From Tab. \ref{tab:results}, we see that $\phi$ achieves the best results, all the measures have positive degrees of correlation with $\gamma$ and small $p$-values (hence statistical significant).

\textbf{Mackey-Glass.}
The input signal in this test is generated by the Mackey-Glass (MG) system, described by the following differential equation:
\begin{equation}
\label{eq:MG}
\frac{dx}{dk} = \frac{\alpha x(k-\tau_{\mathrm{MG}})}{1+ x(k-\tau_{\mathrm{MG}})^{10}} - \beta x(k).
\end{equation}
We generated a time series using $\tau_{\mathrm{MG}} = 17, \alpha = 0.2, \beta = 0.1$, initial condition $x(0)=1.2$, 0.1 as integration step and we trained the system to predict $\tau_f=6$ step ahead.
As we can see from Fig. \ref{fig:MG_task} and the results in the table, for this test both $\phi$ and $\lambda$ provide much better results than $\eta$ for identifying the optimal configuration. Notably, the correlation between $\gamma$ and $\eta$ has a $p$-value beyond the confidence level $0.05$, suggesting that correlations are not different from zero.

\textbf{NARMA.}
This task, originally proposed in \cite{jaeger2002adaptive}, consists in modeling the output of the following order-$r$ system:
\begin{align}
\label{eq:narma}
&\mathbf{y}[k + 1] =\\
\nonumber&0.3\mathbf{y}[k] + 0.05\mathbf{y}[k]\left(\sum \limits_{i=0}^{r-1} \mathbf{y}[k - i]\right) + 1.5\mathbf{x}[k - r]\mathbf{x}[k] + 0.1,
\end{align}
being $\mathbf{x}[k]$ an i.i.d. uniform noise in $[0, 1]$. According to the results shown in Fig. \ref{fig:NARMA_task} and Tab. \ref{tab:results}, in this case $\phi$ and $\eta$ perform significantly better than $\lambda$ for identifying the critical region. If fact, the correlation between $\gamma$ and $\lambda$ is low and not statistically significant. Even in this case, the best results in terms of correlation are achieved by $\phi$.

\bgroup
\def\arraystretch{1.3} 
\setlength\tabcolsep{0.2em} 
\begin{table}
\scriptsize
\caption{Correlations between the regions where FIM determinant is maximized ($\phi$), MLLE crosses zero ($\lambda$), mSVJ is maximized ($\eta$) and performances are maximized ($\gamma$/MC). Best results are shown in bold, $p$-values are reported in brackets.}
\vspace{0.1cm}
\begin{tabular}{lccc}
\cmidrule[1.5pt]{1-4}
\textbf{Test} & \textbf{Corr ($\boldsymbol{\phi}$ , $\boldsymbol{\gamma}$/MC)} & \textbf{Corr ($\boldsymbol{\lambda}$ , $\boldsymbol{\gamma}$/MC)} &   \textbf{Corr ($\boldsymbol{\eta}$ , $\boldsymbol{\gamma}$/MC)} \\
\hline
MC 			& 0.75 (1e-5) 		& \textbf{0.81} (1e-8) 	& 0.65 (1e-4) \\
Predict -- SIN 		& \textbf{0.58} (0.02) & 0.52 (1e-3) 		& 0.56 (1e-3) \\
Predict -- MG 		& \textbf{0.71} (1e-5)	& 0.66 (1e-4)		& 0.38 (0.06) \\
Predict -- NARMA 	& \textbf{0.52} (0.01) 	& 0.25 (0.22) 		& 0.48 (0.02) \\
Predict -- D4D 	& \textbf{0.63} (1e-4)	& 0.34 (0.09)		& 0.54 (0.01) \\
\cmidrule[1.5pt]{1-4}
\end{tabular}
\label{tab:results}
\end{table}
\egroup

\subsection{Prediction of mobile traffic load time series}
\label{sec:orange_call_load}

Here, we analyze time series of data related to nationwide mobile telephone loads.
Such time series have been generated from the data collected in the Orange telephone dataset, published in the Data for Development (D4D) challenge \cite{DBLP:journals/corr/abs-1210-0137}.
D4D is an open collection of call data records, containing anonymized events of Orange's mobile phone users in Ivory Coast, Africa.
More detailed information on the challenge is available on the related website\footnote{http://www.d4d.orange.com}.
The dataset considered here span from December 1, 2011 to April 28, 2012. It includes antenna-to-antenna traffic on an hourly basis, relative to mobile phone calls and SMS.
Each record in the dataset has the following structure: $\langle \mathrm{DateTime}, \mathrm{ID}_\mathrm{a}, \mathrm{ID}_\mathrm{b}, \mathrm{NumCalls}, \mathrm{TotTime} \rangle$.
$\mathrm{DateTime}$ is the time (with hourly resolution) and date when an activity between the two antennas $a$ and $b$ has been registered; $\mathrm{ID}_\mathrm{a}$ and $\mathrm{ID}_\mathrm{b}$ are the identifiers of the transmitting and receiving antenna, respectively; $\mathrm{NumCalls}$ is the number of calls started from $a$ and received by $b$ in the time interval under consideration; finally $\mathrm{TotTime}$ is the sum of the durations (in seconds) of all calls recorded in the interval.
We selected a specific antenna and retrieved from the dataset all those records relative to the activity involving that antenna.
We have accordingly generated the following 7 distinct time series:
\begin{itemize}
  \item[--] \texttt{ts1}: constant input (a time series with all values set to 1). This is a standard practice in prediction with neural networks, since a constant input acts as a bias for the individual neurons of the network \cite{jaeger2002adaptive};
  \item[--] \texttt{ts2}: number of incoming calls in the area covered by the antenna;
  \item[--] \texttt{ts3}: volume in minutes of the incoming calls in the area covered by the antenna;
  \item[--] \texttt{ts4}: number of outgoing calls in the area covered by the antenna;
  \item[--] \texttt{ts5}: volume in minutes of the outgoing calls in the area covered by the antenna;
  \item[--] \texttt{ts6}: hour of the day when the telephone activity was registered;
  \item[--] \texttt{ts7}: day of the week when the telephone activity was registered.
\end{itemize}
All these 7 time series are fed as input to an ESN, while we predict the values relative only to \texttt{ts2}.

\begin{figure}[ht!]
\centering
\includegraphics[scale=0.5]{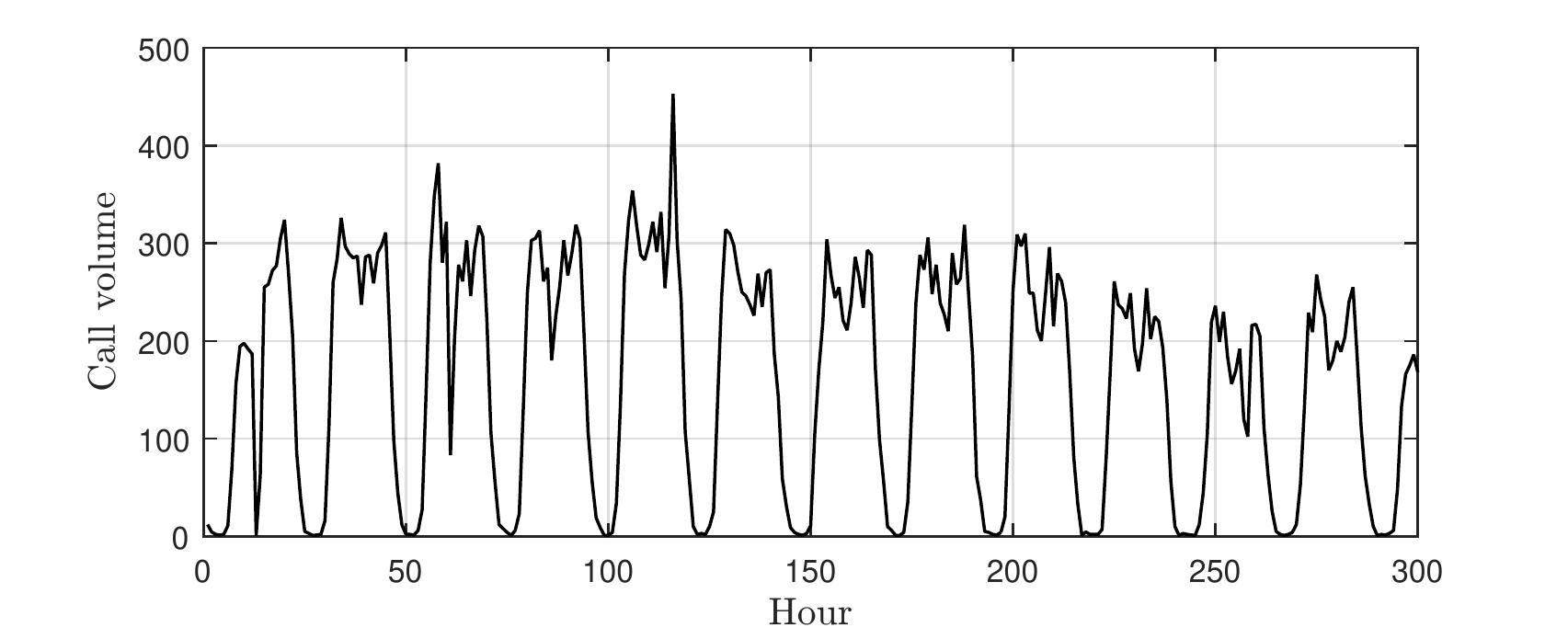}
\caption{The load profile of \texttt{ts2} for the first 300 time intervals.}
\label{fig:callVolume}
\end{figure}

The dataset contains a small percentage of missing values; they appear when there is no outgoing and/or incoming telephone activities for the monitored antenna at a given hour. They are replaced by 0s, to guarantee each time series to have the same length. Corrupted data are relative to periods where the telephone activity is not correctly registered and they are marked with a ``-1'' in the dataset.
They have been replaced with the average value of the corresponding periods (i.e., same weekday and hour of the day) in two adjacent weeks, according to \cite{shen2005analysis}.
All data have been standardized by a z-score transformation prior to processing. This is successively reversed when the forecast must be evaluated.
In Fig. \ref{fig:callVolume} we show the profile of \texttt{ts2} relatively to the load in the first 300 time intervals (corresponding to 1 hour of activity).

D4D time series have been previously studied in \cite{bianchi2015prediction}, where ESN and other standard methods (ARIMA and Triple Exponential Smoothing) were adopted to perform both 1-step and 24-steps ahead predictions.
In \cite{bianchi2015prediction}, hyperparameters were tuned using a genetic algorithm (GA) optimization scheme and different optimization procedures were evaluated for training the readout: least-square regression; elastic net penalty; linear and nonlinear $\nu$-SVR.
It was shown that ESN achieved higher prediction accuracy with respect to the other forecast methods, especially when using non-linear $\nu$-SVR, at the expense of a much higher computational cost.
It is worth mentioning that, when GAs are used in a supervised cross-validation scheme, they yield a single solution that is black-box and does not follow a mathematically motivated criterion to determine the edge of criticality.
Additionally, a non-linear model such as $\nu$-SVR, must be trained at each iteration of the optimization procedure, with consequent huge increase in time complexity.

We focus again on the analysis of the hyperparameters $\theta_{SR}$, $\theta_{IS}$, and $\theta_{RC}$, while for the remaining ones we adopt the optimal configuration found in \cite{bianchi2015prediction}.
In particular, we set $N_r = 680$ and the regularization parameter in the ridge regression equal to $0.04$. 
Now, the forecast step $\tau_f$ is set to 1, i.e., we predict the telephonic load of the next hour.
Note that this differs from the other tasks previously considered, where $\tau_f$ was set equal to the first zero of the autocorrelation function.

In Fig. \ref{fig:ORANGE_task}, we show the results of $\gamma$ (prediction accuracy on \texttt{ts2}) with respect to $\phi$, $\lambda$, and $\eta$.
According to Tab. \ref{tab:results}, even in this case $\phi$ gives rise to the manifold with higher correlation with $\gamma$; $\eta$ produces lower yet positive and statistically significant correlations; finally, $\lambda$ achieves the worst results.
Interestingly, by using the FIM-based criterion we find a critical region in the three-dimensional ESN hyperparameter space containing the optimal values for $\theta_{SR}$ and $\theta_{IS}$, which are reported in \cite{bianchi2015prediction}.
In fact, the quantized area centered in $\{ \theta_{SR}=1, \theta_{IS}=0.35, \theta_{RC}=0.55 \}$, belongs to the ESN critical region, according to the FIM-based criterion.
Such region contains also the values $\theta_{SR} = 0.98$ and $\theta_{IS} = 0.33$, which were identified as optimal in \cite{bianchi2015prediction}.
Instead, for what concerns the sparsity of reservoir connectivity $\theta_{RC}$, the upperbound considered in the GA optimization for this value was set to 0.4. Accordingly, the optimal configuration with $\theta_{RC}=0.55$ could not be obtained. This represents a good example of how the proposed method provides a more flexible approach for tuning the network hyperparameters. In fact, while several cross-validation methods (using a GA is just an example) treat the model as a black-box, the framework we developed allows to analyze the dynamics of the systems and to visualize the critical regions in the process of optimizing the network. For example, it is possible to assess if the critical region is too close to one of the bounds considered for a given hyperparameter, allowing to redefine the bounds accordingly.


\section{Conclusions and future directions}
\label{sec:conclusions}

Echo state networks, as a class of networks in reservoir computing, offer a compromise between training time and network performance in terms of prediction error and short-term memory capacity.
Experiments showed that such networks operate more efficiently when configured on the so-called edge of criticality, a region in hyperparameter space separating ordered and chaotic regimes.
Hyperparameters (indirectly) affecting the behavior of the network are hence tuned according to some criterion.
In this paper, we proposed a principled approach for configuring an echo state network on the edge of criticality.
The proposed method is completely unsupervised and is based on the interplay between the theory of continuous phase transitions and Fisher information.
In fact, it is possible to prove that Fisher information diverges on the critical region and hence can be used to determine the onset of criticality.
Nonetheless, Fisher information presumes analytic knowledge of the parametric distribution describing the system/network; in addition its computation is known to be difficult and prone to numerical errors.
In order to deal with these issues, here we have followed an ensemble estimation approach based on a recently proposed non-parametric Fisher information matrix estimator. Such an estimator is applicable to high-dimensional densities, since it operates by means of a graph-based representation of the data.
This last aspect is very important in our case, since we analyze the network though a multivariate sequence of reservoir neuron activations.

We evaluated the proposed method on well-known benchmarks as well as on a real-world application involving telephone call load prediction.
The benchmarks taken into account were conceived to evaluate both the short-term memory capacity (in terms of the squared correlation between past inputs and network outputs) and the prediction accuracy (in terms of normalized root mean square error).
In order to compare our method with other unsupervised approaches, we have taken into account (i) a criterion based on the sign of the maximum local Lyapunov exponent computed on the activations and (ii) a criterion based on the maximum value of the minimum singular value of the Jacobian matrix of the reservoir.
Results showed that the proposed method based on Fisher information is more accurate than those two unsupervised methods (on both the benchmarks and the real-world application) in determining critical ESN hyperparameter configurations in terms of accuracy. However, test on memory capacity showed that Lyapunov exponent provides better results, although differences are not as high as in the accuracy case.

The methodology proposed here offers a sound and appealing solution to determine the onset of criticality in echo state networks, with potential extension to other families of recurrent neural networks.
Nonetheless, our contribution comes with some technical difficulties that we have only partially solved so far.
First of all, potential non-stationarities and dependencies of the neuron activations might affect the estimation outcomes.
Here, we have addressed this issue by following an ensemble approach to estimate the Fisher information matrix. However, other approaches might be considered in the future, for instance by following a window-based analysis scheme.
Second, the non-parametric Fisher information matrix estimator we used requires to set a parameter controlling the magnitude of the perturbations. This parameter turned out to be very sensitive and difficult to determine in practice, hence posing some technical limitations when trying to automatize the procedure. Such issues will be object of future research.

There are many possible routes that we intend to follow in future research studies. Among the many, we believe it is worth focusing on (i) how to enable and control the output feedback connections and (ii) the application of the proposed method as a unsupervised learning method for other types of recurrent neural networks.


\section*{Acknowledgements}

We wish to thank Prof. Visar Berisha and Prof. Alfred Olivier Hero III for providing us with part of the code required to implement the non-parametric Fisher information matrix estimator used in this work.

\begin{appendices}
\section{Proposed formulation of the semidefinite constraint}
\label{sec:constraint_implementation}

Here we provide the details of the formulation in matrix form of the $\mathrm{mat}(\cdot)$ operator in Eq. \ref{eq:PSD_opt}.
This is a necessary step to implement the semidefinite constraint in matrix form.

First, we express the constraint with $\mathrm{mat}(\cdot)$ using the inverse operator, $\mathrm{vec}(\cdot)$, which transforms a matrix into its vector representation.
A matrix $\mathbf{F} \in \mathbb{R}^{m \times n}$ is converted into the vector representation as follows:
\begin{equation}
\label{eq:vectorization}
\mathbf{F}_{\mathrm{vec}} = \sum \limits_{i=1}^n \mathbf{B}_i\mathbf{F}\mathbf{E}_i,
\end{equation}
where $\mathbf{E}_i$ is the $i$-th canonical basis vector of an $n$-dimensional Euclidean space, i.e., $\mathbf{E}_i = \left[ 0, \ldots,0,1,0,\ldots,0\right]^T$ has a 1 in the $i$-th position and 0 elsewhere.
$\mathbf{B}_i$ is a $(mn) \times m$ block matrix defined as a stack of $n$ blocks, which are defined as $m \times m$ zero-matrix with the exception of the $i$-th block, which is the identity matrix:
\begin{align}
\mathbf{B}_i = \left[ \mathbf{0}_{m\times m}, \cdots, \mathbf{0}_{m\times m}, \mathbb{I}_{m\times m}, \mathbf{0}_{m\times m}, \cdots, \mathbf{0}_{m\times m} \right]^T.
\end{align}

Notice that, in our case, $m = n = d$, where $d$ is the number of ESN hyperparameters taken into account. To convert the half-vector representation $\mathbf{F}_{\mathrm{hvec}}$ in Eq.  \ref{eq:PSD_opt} into the vector form $\mathbf{F}_{\mathrm{vec}}$, we rely on the following expression:
\begin{equation}
\mathbf{D}\left(\mathbf{S}\mathbf{F}_{\mathrm{hvec}}\right) = \mathbf{F}_{\mathrm{vec}},
\end{equation}
where $\mathbf{D}$ and $\mathbf{S}$ are the multiplication and the shuffling matrices, respectively.
These matrices cannot be expressed in closed-form \cite{magnus1995matrix}.
Therefore, in the following we provide the pseudo-code of Algorithms \ref{alg:duplication_matrix} and \ref{alg:shuffling_matrix} that implement them.
\begin{algorithm}\footnotesize
\caption{Duplication matrix computation.}
\label{alg:duplication_matrix}
\begin{algorithmic}[1]
\REQUIRE Dimensionality $d$ of the hyperparameter space
\ENSURE Duplication matrix, $\mathbf{D}$
\STATE $\mathbf{D} = \mathbb{I}_{d^2 \times d^2} = \left[\mathbf{d}_1, \ldots, \mathbf{d}_{d^2}\right]^T$
\STATE $\gamma = \delta = \emptyset$
\FOR{$i = 1, \ldots, d-1$}
        \STATE $\gamma \cup \left\{ i + di, \ldots, i + d(d-1) \right\}$
        \STATE $\delta \cup \left\{ i + d(i-1) + 1, \ldots, i + d(i-1) + d-i \right\}$
\ENDFOR
\FOR{$i = 1, \ldots, d(d-1)/2$}
        \STATE $\mathbf{d}_{\delta(i)} = \mathbf{d}_{\gamma(i)} + \mathbf{d}_{\delta(i)}$
\ENDFOR
\FOR{$i = 1, \ldots, d(d-1)/2$}
        \STATE remove $\mathbf{d}_{\gamma(i)}$ from $\mathbf{D}$
\ENDFOR
\end{algorithmic}
\end{algorithm}

\begin{algorithm}\footnotesize
\caption{Shuffling matrix computation.}
\label{alg:shuffling_matrix}
\begin{algorithmic}[1]
\REQUIRE Dimensionality $d$ of the hyperparameter space
\ENSURE Shuffling matrix, $\mathbf{S}$
\STATE $\mathbf{S} = \mathbf{0}_{d(d+1)/2 \times d(d+1)/2} = \left[\mathbf{s}_1, \ldots, \mathbf{s}_{d(d+1)/2}\right]^T$
\STATE $\mathbf{I} = \mathbb{I}_{d(d+1)/2 \times d(d+1)/2} = \left[\mathbf{i}_1, \ldots, \mathbf{i}_{d(d+1)/2}\right]^T$
\STATE $\mathbf{s}_1 = \mathbf{i}_1$
\FOR{$j = 2, \ldots, d$}
        \STATE $\gamma = 1+ d(j-1) - (j-1) (j-2)/2$
        \STATE $\mathbf{s}_j = \mathbf{i}_{\gamma}$
        \STATE remove $\mathbf{i}_{\gamma}$ from $\mathcal{I}$
\ENDFOR
\FOR{$j = d+1, \ldots, d(d+1)/2$}
        \STATE $\mathbf{s}_j = \mathbf{i}_j$
\ENDFOR
\end{algorithmic}
\end{algorithm}

Hence, the optimization problem in (\ref{eq:PSD_opt}) is formalized as:
\begin{equation}
\label{eq:PSD_opt2}
\begin{aligned}
& \underset{\mathbf{F}_{\mathrm{hvec}}}{\text{minimize}}
& & \lVert \mathbf{R} \mathbf{F}_{\mathrm{hvec}} - \mathbf{v}_{\boldsymbol{\theta}} \rVert^{2} \\
& \text{subject to} & & \mathbf{F}_{\mathrm{hvec}}(i) = \mathbf{\hat{F}}_{\mathrm{hvec}}(i),\ i \in \{1, \ldots, d\}, \\
& & & \mathbf{D}\left(\mathbf{S}\mathbf{F}_{\mathrm{hvec}}\right) = \mathbf{F}_{\mathrm{vec}}, \\
& & & \mathbf{F}_{\mathrm{vec}} = \sum \limits_{i=1}^d \mathbf{B}_i \mathbf{F} \mathbf{E}_i, \\
& & & \mathbf{F} \succeq \mathbf{0}_{d\times d}.
\end{aligned}
\end{equation}

\end{appendices}

\bibliographystyle{./IEEEtran}
\bibliography{Bibliography}

\end{document}